\documentclass[twocolumn,secnumarabic,amssymb,amsmath,nobibnotes, aps, pra]{revtex4-1}
\usepackage{graphicx}
\usepackage{dcolumn}
\usepackage{bm}
\usepackage[dvipdfmx]{hyperref}
\usepackage{subfigure}

\begin{document}

\title{Multipartite entanglement, quantum coherence and quantum  criticality in triangular and  Sierpi\'nski fractal lattices}%

\author{Jun-Qing Cheng}%
\author{Jing-Bo Xu}%
\email[Emali: ]{xujb@zju.edu.cn}
\affiliation{Zhejiang Institute of Modern Physics and Physics Department, Zhejiang University, Hangzhou 310027, China}
\date{\today}%
\begin{abstract}
We investigate the quantum phase transitions of the  transverse-field quantum Ising model on the triangular lattice and Sierpi\'nski  fractal  lattices by employing the multipartite entanglement and quantum coherence along with the quantum renormalization group method. It is shown that the quantum criticalities of these high-dimensional models closely relate to the behaviors of the multipartite entanglement and quantum coherence. As the thermodynamic limit is approached, the first derivatives of the multipartite entanglement and quantum coherence exhibit singular
 behaviors,  and the consistent finite-size scaling behaviors for each lattice  are also obtained from the first derivatives. The multipartite entanglement and quantum coherence are demonstrated to be  good indicators for detecting the quantum phase transitions in the triangular lattice  and Sierpi\'nski fractal lattices. Furthermore, the dimensions determine the relations between the critical exponents and the correlation length exponents for these lattices.

 \begin{description}
\item[PACS numbers]
\pacs{}05.30.Rt, 03.67.Mn, 75.10.Pq, 64.60.al
\end{description}
\end{abstract}

\maketitle

\section{Introduction}

Quantum phase transitions (QPTs) are  notable manifestations of quantum many-body systems at absolute zero temperature where the quantum fluctuations play  a dominant role and no thermal fluctuations exist  \cite{Sachdev1999}. QPTs can be achieved by changing the  parameters of  Hamiltonian, such as an external magnetic field or the coupling constant. As a control parameter is varied through a critical value,  the ground state of a system suffers an abrupt change mapped to a  variation in the system's properties.  How to reveal and characterize the critical phenomenons of quantum many-body systems is an important task  and becomes a hot topic in condensed-matter physics.  Traditional  methods mainly focus on the identification of the order parameters and the pattern of symmetry breaking. Recent developments in quantum information theory \cite{Nielson2000Quantum} have provided  some insights into the QPTs. Specifically,  the quantum entanglement has been successfully used as an effective tool to reveal the QPTs without any prior knowledge of the
order parameter \cite{Osborne2002,Nature2002Scaling,Wu2004Quantum,Amico2008}. Since  the concept of renormalization was introduced from the quantum field theory to quantum statistical physics \cite{Wilson1975The}, many progresses have been made in the research of QPTs. As a variant of renormalization group at zero temperature, quantum renormalization group (QRG) is a  tractable method for studying the criticalities of  one-dimensional \cite{jafari2007phase,Ma2011Entanglement,PhysRevA.86.042102,Efrati2014Real} and two-dimensional \cite{Usman2015Quantum,PhysRevA.95} many-body  systems. This method can be used to evaluate  the quantum critical points and scaling behaviors analytically, but has difficulties in quantitative estimation for the    transverse-field Ising model \cite{PhysRevB.18.3568,PhysRevB.19.4653}. Recently,  it has been shown that a  novel   renormalization group (RG) map can not only  be
used to accurately examine  the critical behavior of the one-dimensional quantum transverse-field Ising model, and also be  used to predict the critical behaviors of the  higher-dimensional models  \cite{Miyazaki2011Real,kubica2014precise}. In particular, there have been efforts to study the quantum Ising models on fractal lattices \cite{Yi2013Critical,kubica2014precise,PhysRevE.91.012118} which
were not clear before. Fractals are self-similar structures  in noninteger dimensions and have both aesthetic and scientific interests. They have been used to interpolate between integer dimensional regular lattices and construct the networks  for quantum computation and communication \cite{markham2013,Michael2016}. The quantum criticalities of fractal lattices attract our attention.

On the other hand, the  entanglement in the ground
state of a many-body system can be utilized as a resource for quantum
technologies \cite{Amico2008}. The multipartite entanglement offers significant advantages in quantum tasks compared with bipartite entanglement.  For example, it is the main  ingredient in measurement-based  quantum computation \cite{ Browne2009} and  various quantum communication protocols \cite{RevModPhys.74.145,PhysRevLett.86.4431,PhysRevLett.86.5188}. Therefore, the entanglement quantification  of multipartite quantum states is necessary and essential in quantum information science. The monogamy of entanglement is one of most important properties in many-body quantum systems \cite{Horodecki2007}, and  can be used to characterize the entanglement structure. It has been recently discovered that the squared entanglement of
formation obeys the monogamy inequality in an arbitrary $N$-qubit mixed state, and a relevant multipartite entanglement indicator is proposed \cite{Bai2014General}. The multipartite entanglement
provides a global view and more physical insights into the characters of a many-body
system, and it may have some advantages over bipartite entanglement to reveal the QPTs.    Furthermore, the quantum coherence, which arises from  the quantum superposition principle,  plays a very important role in the fields of quantum optics \cite{Louisell1973Quantum} and quantum information \cite{Nielson2000Quantum}.  However, there has been no well-accepted efficient method for measuring the quantum coherence until very recently.   A rigorous theoretical framework for quantifying the quantum coherence and the necessary constraints for the quantifier have been proposed \cite{PRLcoherence}.  It is interesting to  do some research about the multipartite entanglement and  quantum coherence in the  QPTs of high-dimensional many-body systems.

These developments on QPTs, QRG method, multipartite entanglement and quantum coherence motivate us to consider the following questions: How do the multipartite entanglement and quantum coherence behave in the QPTs of high-dimensional models? Can the multipartite entanglement and quantum coherence  be used to indicate the  QPTs of the transverse-field quantum Ising models on the  fractal  lattices? If we can apply the QRG approach  to find  the finite-size scaling behaviors  proposed in Ref. \cite{Nature2002Scaling} for the cases of  fractal  lattices?  Are the critical exponents of
multipartite entanglement consistent with the ones of quantum coherence for the same lattice?  What are the relations between the critical exponents and  correlation length exponents for high-dimensional systems?
In this paper, we investigate the performances of multipartite entanglement and quantum coherence in the QPTs for transverse-field quantum Ising model on the triangular lattice and Sierpi\'nski  fractal  lattices by employing the QRG method.  It is found that the quantum criticalities of these models closely relate to the behaviors of the multipartite entanglement and  quantum coherence.  The singularities   for each lattice  are observed from the first derivatives of the multipartite entanglement and quantum coherence.
The scaling behaviors as  introduced in Ref. \cite{Nature2002Scaling} are obtained for these lattices,  especially the  ones which describe how the critical points  are touched  as the  thermodynamic limit is approached haven't been discussed before.
It is also shown that the multipartite entanglement and quantum coherence obey the universal finite-size scaling laws for the same lattice. Furthermore,
the  dimensions of lattices  play the
decisive roles on the relations between the critical exponents and  correlation length exponents.  The multipartite entanglement and quantum coherence are proven to be  good indicators to detect the QPTs of the transverse-field quantum Ising model on the high-dimensional lattices, such as the triangular and Sierpi\'nski fractal lattices.

This paper is organized as follows.  In Sec. \ref{section2}, we  study the QPTs of lattices by employing multipartite entanglement  along with the QRG method.  In Sec. \ref{section3}, we investigate  the quantum coherence   and the QPTs of lattices by using the QRG method.
 Finally, the conclusions are drawn in Sec. \ref{section4}.

\section{\label{section2}Multipartite entanglement and Quantum phase transitions in lattices}

\begin{figure}[ht]
	\centering
	\subfigure[]{ \label{figure1a} \includegraphics[width=7cm]{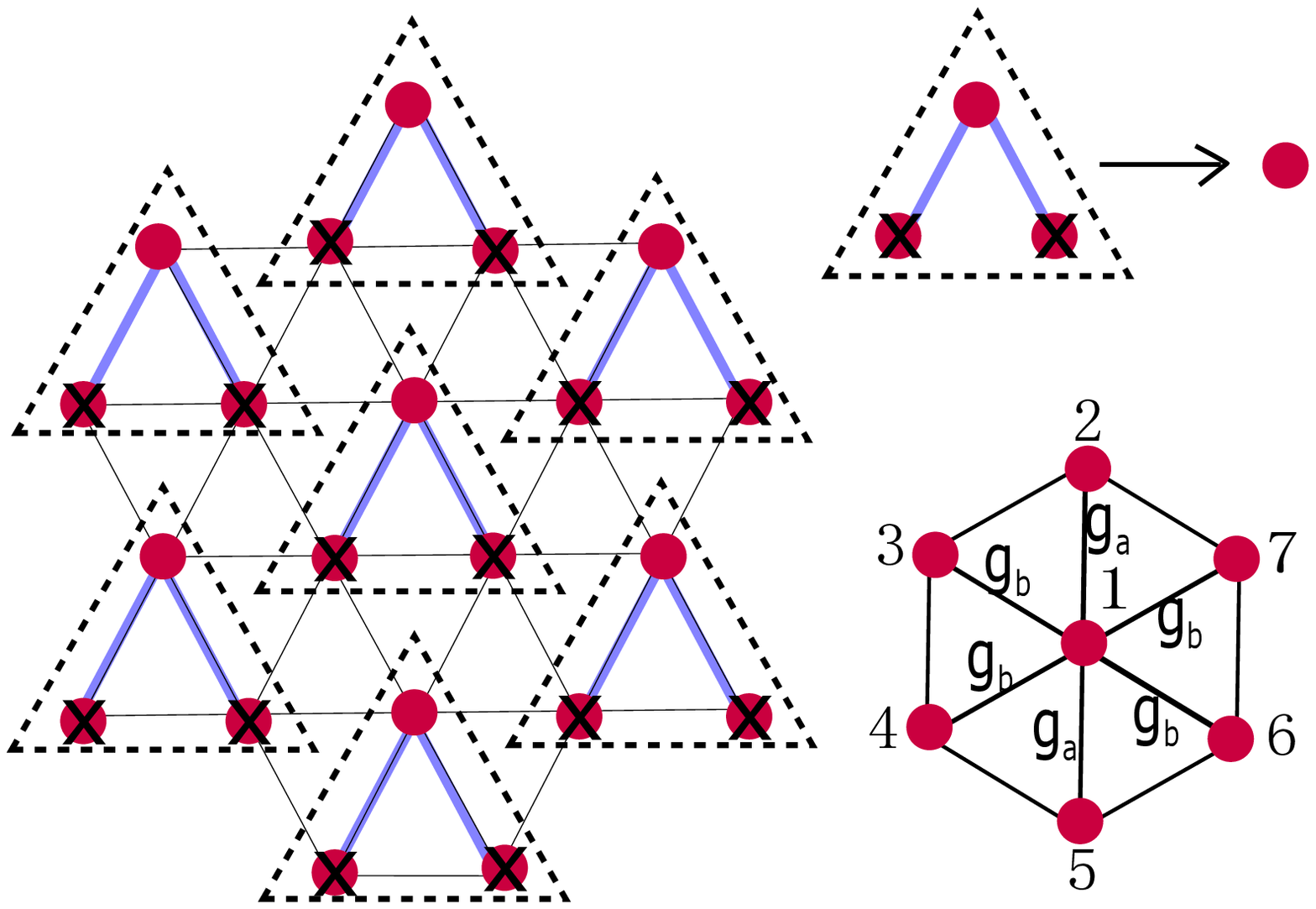}}
    \subfigure[]{\label{figure1b}
\includegraphics[width=4cm]{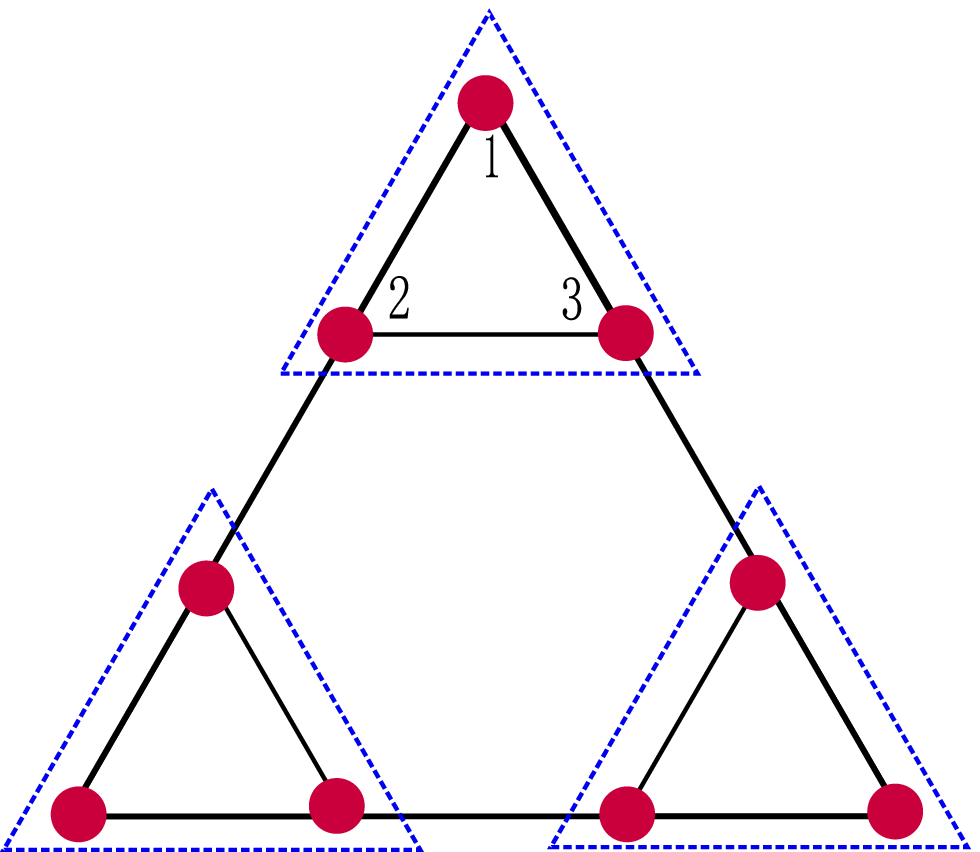}}
	\subfigure[]{\label{figure1c} \includegraphics[width=4cm]{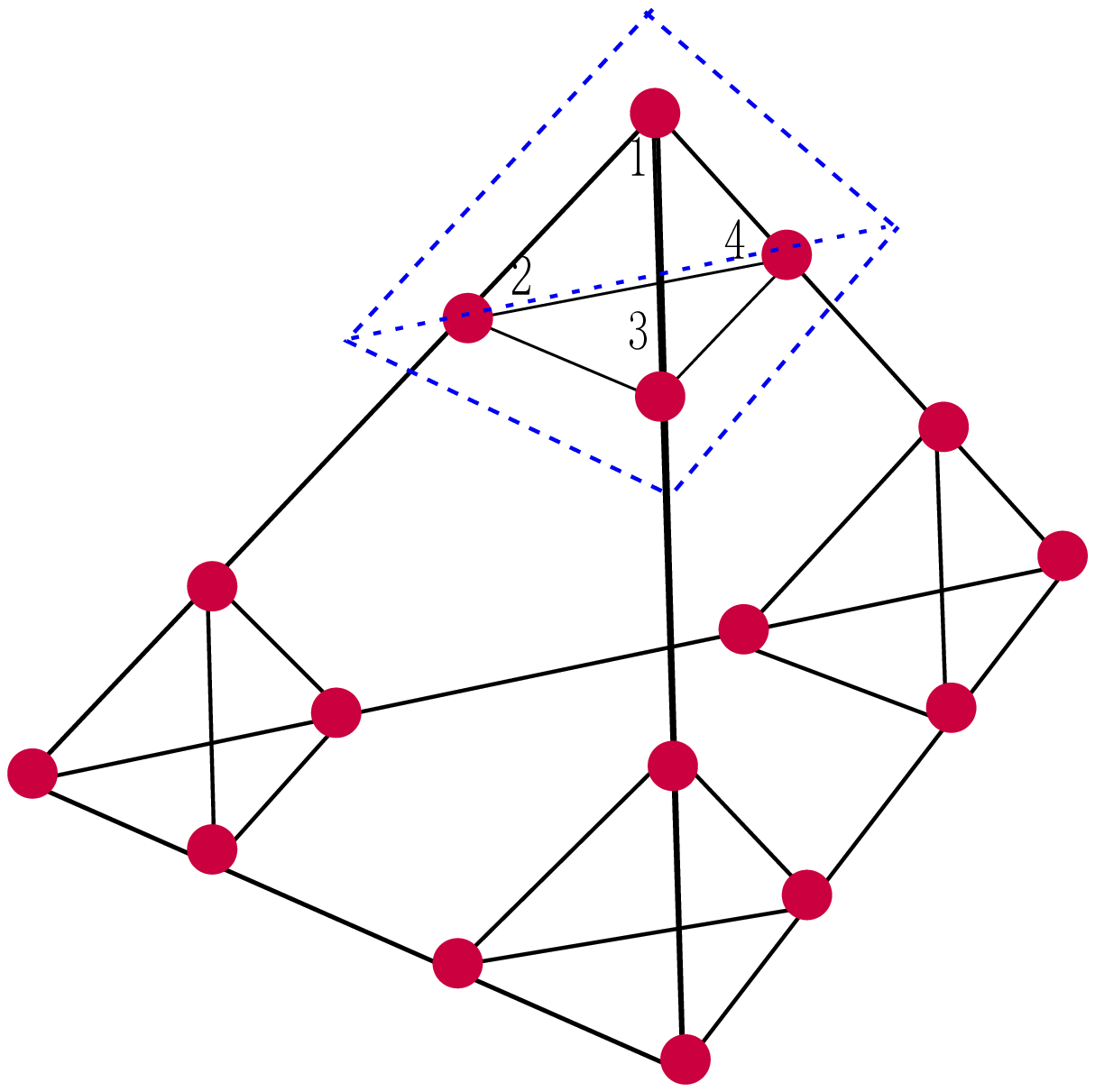}}
\caption{Schematic illustration of QRG transformation for the (a)
		triangular lattice and (b)(c) Sierpi\'nski fractal lattices.}
	\label{figure1}
\end{figure}

We consider a set of localized spin-$1/2$ particles in the triangular lattice  or Sierpi\'nski fractal lattices coupled through exchange
interaction $J$ and subject to an external magnetic field of strength $h$. The Hamiltonians for such transverse-field quantum Ising models  are given by
\begin{equation}
H=-J \sum_{\left\langle i,j\right\rangle }\sigma_i^z \sigma_j^z -h\sum_{i}\sigma_i^x
\end{equation}
where  $\sigma_i^\alpha (\alpha=x,z)$  are the standard spin-$1/2$ Pauli operators at the site $i$. The sums are over all the nearest neighbor pairs and over all sites, respectively. We mainly focus on the ferromagnetic interactions $J> 0$ and the transverse field $h\geqslant 0$.  In this work,  three kinds of  lattices as shown in Fig. \ref{figure1} are considered, which are the  triangular lattice and  Sierpi\'nski  fractal  lattices, respectively.  For simplicity, the   exchange interaction normalized to  the transverse field strength $g=J/h$ is applied during our investigation.

Typically, it is not easy to obtain the analytical solutions of these high-dimensional systems. Even if the numerical method such as Monte-Carlo simulation is applicable \cite{Hasenbusch2010A}, the calculation is computationally expensive. The QRG method is a  analytical treatment for studying the QPTs, especially has advantages in estimating the quantum critical points and scaling behaviors.
The main idea of QRG method is to eliminate or thin the degrees of freedom of the many-body systems through a recursive procedure until a  tractable situation is reached.
According to the Kadanoff's block method \cite{Efrati2014Real,jafari2007phase,PhysRevA.86.042102}, a spin chain can be split into blocks, which means that the Hamiltonian is decomposed into the block Hamiltonian and interacting (interblock) Hamiltonian. The low-lying eigenstates of each block Hamiltonian  are applied to construct the basis for renormalized Hilbert space. In this way, the full Hamiltonian is projected onto the renormalized space to achieve an effective Hamiltonian with structural similarity to the original one. As long as the thermodynamic limit is touched by increasing the RG iterations,  the global properties of the system  can be captured.

In the  novel   QRG method, for the purpose of preserving the symmetry of  system and the  structural similarity of  Hamiltonian, not all the terms inside a block is included in the QRG transformation \cite{kubica2014precise}, which is used to
significantly improve the  estimation precision about the critical point. The  procedure of QRG transformation for the triangular lattice  is shown in Fig. \ref{figure1a}.
 The entire system is covered by blocks of three sites which are renormalized to be a single one.
After performing the renormalization,  two types of coupling strengths $g_a$ and $g_b$ replace the original one $g$, which means that the coupling strength becomes highly anisotropic.  In Ref. \cite{kubica2014precise}, the authors have proposed that the renormalized coupling strength for  triangular lattice should be a geometric mean of all coupling strengths and choose the hexagon  consisting of  seven sites as a basic  cluster.  Then the renormalized coupling strength  can be obtained by
\begin{equation}
\label{RG equation1}
g'_\mathrm{t}=g_a^{2/6} g_b^{4/6}=2^{1/3} g^2 (1+g^2)^{1/6} (g+\sqrt{1+g^2})^{2/3}.
\end{equation}
In this way, the triangular lattice with $ 7\times (\lambda^n)^d $ sites can be effectively represented by a seven-site cluster after completing the $n$th RG iteration step, where $d=2$ is the dimension of triangular lattice and $\lambda= \sqrt{3}$ is the scale of the length of  the side for  each RG iteration. The   critical point $g_\mathrm{c}^t$  corresponding to the nontrivial fixed point  is obtained by solving $g'=g$, i.e., $g_\mathrm{c}^\mathrm{t}\approx 0.539$. Similarly, we also study the  transverse-field quantum Ising model on the Sierpi\'nski fractal lattices with Hausdorff dimension $d_\mathrm{H}=\log(\kappa+1)/\log2$ where $\kappa=2$ or $3$ is the spatial dimension.
The  procedures of QRG transformation for the Sierpi\'nski triangular lattice ($d_\mathrm{H}=1.585$) and Sierpi\'nski pyramid lattice ($d_\mathrm{H}=2$) are depicted in  Fig. \ref{figure1b} and Fig. \ref{figure1c}, respectively \cite{kubica2014precise}.  It can be observed that the basic cluster in the Sierpi\'nski triangular lattice  is  a triangle  containing three sites, and for the Sierpi\'nski pyramid lattice, it is a pyramid containing four sites.  It is not difficult to find that after $n$th RG iterations the  Sierpi\'nski triangular (or pyramid) lattice with $3\times\lambda_{\mathrm{f}}^{1.585n}$ (or $4\times\lambda_{\mathrm{f}}^{2n}$) sites is represented by a three (or four)-site cluster, where $\lambda_\mathrm{f}=2$. Therefore,  the  renormalized coupling strengths for the fractal lattices   can be obtained as follows
\begin{equation}
\label{RG equation2}
g'_\mathrm{f}=g^{(3\kappa+1)/(\kappa+1)}(1+g^2)^{\kappa(\kappa-1)/2(\kappa+1)}.
\end{equation}
The critical points of the Sierpi\'nski triangular lattice and Sierpi\'nski pyramid lattice are given as  $g_\mathrm{c}^\mathrm{St} \approx 0.869$ and $g_\mathrm{c}^\mathrm{Sp} \approx 0.786$, respectively.  Moreover, the correlation length
exponents $\nu$ for the triangular lattice, Sierpi\'nski triangular lattice  and Sierpi\'nski pyramid lattice can be calculated as follows
\begin{eqnarray}
\label{correlation length exponent}
&&\nu_\mathrm{t}^{-1}=\log_{\sqrt{3}} \frac{\mathrm{d}g'_\mathrm{t}}{\mathrm{d}g} |_{g_\mathrm{c}^\mathrm{t}},\nonumber\\
&&\nu_\mathrm{f}^{-1}=\log_2 \frac{\mathrm{d}g'_\mathrm{f}}{\mathrm{d}g} |_{g_\mathrm{c}}.
\end{eqnarray}
The results are  $\nu_\mathrm{t} \simeq 0.630$, $\nu_\mathrm{St} \simeq 0.720$ and $\nu_\mathrm{Sp} \simeq 0.617$, respectively.

Next, we briefly outline the definition  of the monogamy of entanglement and the measure of multipartite entanglement in the present study. For an $N$-qubit system  with state space ${\mathcal{H}_{A_1}} \otimes {\mathcal{H}_{A_2}} \otimes  \cdots \otimes {\mathcal{H}_{A_N}}$, taking the subsystem $A_1$ as a  \textquotedblleft node\textquotedblright \cite{Fanchini2017},
if the entanglement between the particles $A_1$ and $A_2,\cdots,A_N$ satisfies the inequality
\begin{eqnarray}
E^2_{A_1|A_2,\cdots,A_N}\geq E^2_{A_1 A_2}+E^2_{A_1 A_3}+\cdots + E^2_{A_1 A_N},
\end{eqnarray}
with $E_{A_1|A_2,\cdots,A_N}$ quantifying the entanglement in the partition $A_1|A_2,\cdots,A_N$ and $E_{A_1 A_j}$  quantifying the one in the two-qubit system $A_1 A_j$, then the entanglement measure $E$ obeys the monogamous relation \cite{Horodecki2007}.
This monogamy property imposes physical restrictions on unconditional sharability of quantum entanglement between the different parts of a many-body system.  According to the Schmit decomposition \cite{Peres1995},
the subsystem $A_2,\cdots,A_N$ is equal to a logic qubit $A_{2,\cdots,N}$ for an $N$-qubit pure state   $\left| \psi\right\rangle_{A_1 A_2,\cdots,A_N}$. As an example, the entanglement of formation $E_{\rm{f}} (A_1|A_2,\cdots,A_N)$ can be derived by using the   analytical formula for a two-qubit state  $\rho_{AB}$ \cite{Caves2001}
\begin{eqnarray}
\label{bipartite entanglement}
E_{\rm{f}} ({\rho_{AB}})=h \left( \frac{1+\sqrt{1-C^2_{AB}}}{2}\right),
\end{eqnarray}
where $h(x)=-x\log_2{x}-(1-x)\log_2{(1-x)}$ is the binary entropy and $C_{AB}=\max \{  0, \sqrt{\lambda_1}-\sqrt{\lambda_2}-\sqrt{\lambda_3}-\sqrt{\lambda_4} \} $ is the concurrence \cite{Hill1997} with decreasing nonnegative $\lambda_i$s being the eigenvalues of the matrix $\rho_{AB}(\sigma_y\otimes\sigma_y)\rho_{AB}^\ast(\sigma_y\otimes\sigma_y)$.
The  squared  entanglement of formation has been  found to  obey the  monogamy inequality in an arbitrary $N$-qubit mixed state, and a relevant indicator  has been  proposed  to detect the  multiqubit entangled states \cite{Bai2014General}, which reads
\begin{eqnarray}
\tau_{A_1|A_2, \cdots ,A_N} = E_{\rm{f}}^2({\rho _{A_1|A_2, \cdots ,A_N}}) - \sum^N\limits_{j \ne 1} {E_{\rm{f}}^2({\rho _{A_1 A_j}})}
\label{multipartite entanglement}.
\end{eqnarray}
By utilizing this  measure, we can not only  explore the QPTs of many-body systems, but also examine the performance of multipartite entanglement in different phases.

\subsection{triangular lattice}

Now, we investigate the multipartite entanglement  for the transverse-field quantum Ising model  on the triangular lattice by employing the QRG method. Since the basic cluster  contains  seven  sites, as shown in Fig. \ref{figure1a}, we choose the central site (labeled by $1$) as the \textquotedblleft node\textquotedblright \cite{Fanchini2017}, and calculate the seven-partite entanglement $\tau_{1|2, \cdots ,7}$  for studying the performances of multipartite entanglement in the  QPT. The Hamiltonian of basic cluster  can be written as
\begin{equation}
H=-g(\sum_{i=2}^{7}\sigma_1^z\sigma_i^z + \sum_{i=2}^{6}\sigma_i^z\sigma_{i+1}^z +\sigma_2^z\sigma_7^z)-\sum_{i=1}^{7}\sigma_i^x.
\end{equation}
The density matrix is given by $\rho=|\psi_0 \rangle \langle \psi_0|$, where $|\psi_0 \rangle $ is the ground state of the Hamiltonian of the basic cluster. Then we can calculate the seven-partite entanglement according to Eq. \ref{multipartite entanglement}, namely,
  \begin{eqnarray}
 \tau_{1|2, \cdots ,7} = E_{\rm{f}}^2({\rho _{1|2, \cdots ,7}}) - \sum^7\limits_{j \ne 1} {E_{\rm{f}}^2({\rho _{1 j}})}.
 \label{seven-partite}
 \end{eqnarray}

 \begin{figure}[htpb]
 	\centering
 	\includegraphics[width=7cm]{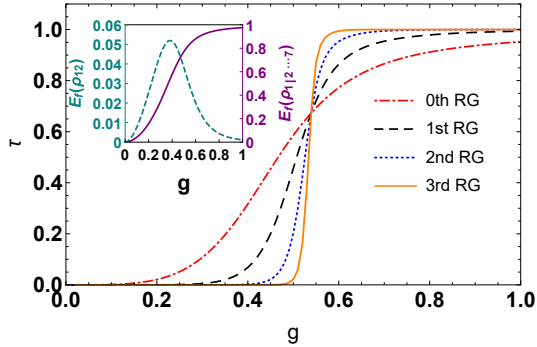}
 	\caption{The evolution of multipartite entanglement $\tau$ versus coupling strength $g$ for different RG iterations on the triangular lattice. The inset depicts the bipartite entanglement $E_{\rm{f}}(\rho_{12})$ (green dashed line) and $E_{\rm{f}}(\rho_{1|2\cdots7})$ (purple solid line) as functions of  coupling strength $g$ for the zeroth RG iteration.}
 	\label{figure2}
 \end{figure}
 \begin{figure}[htpb]
 	\centering
 \includegraphics[width=8cm]{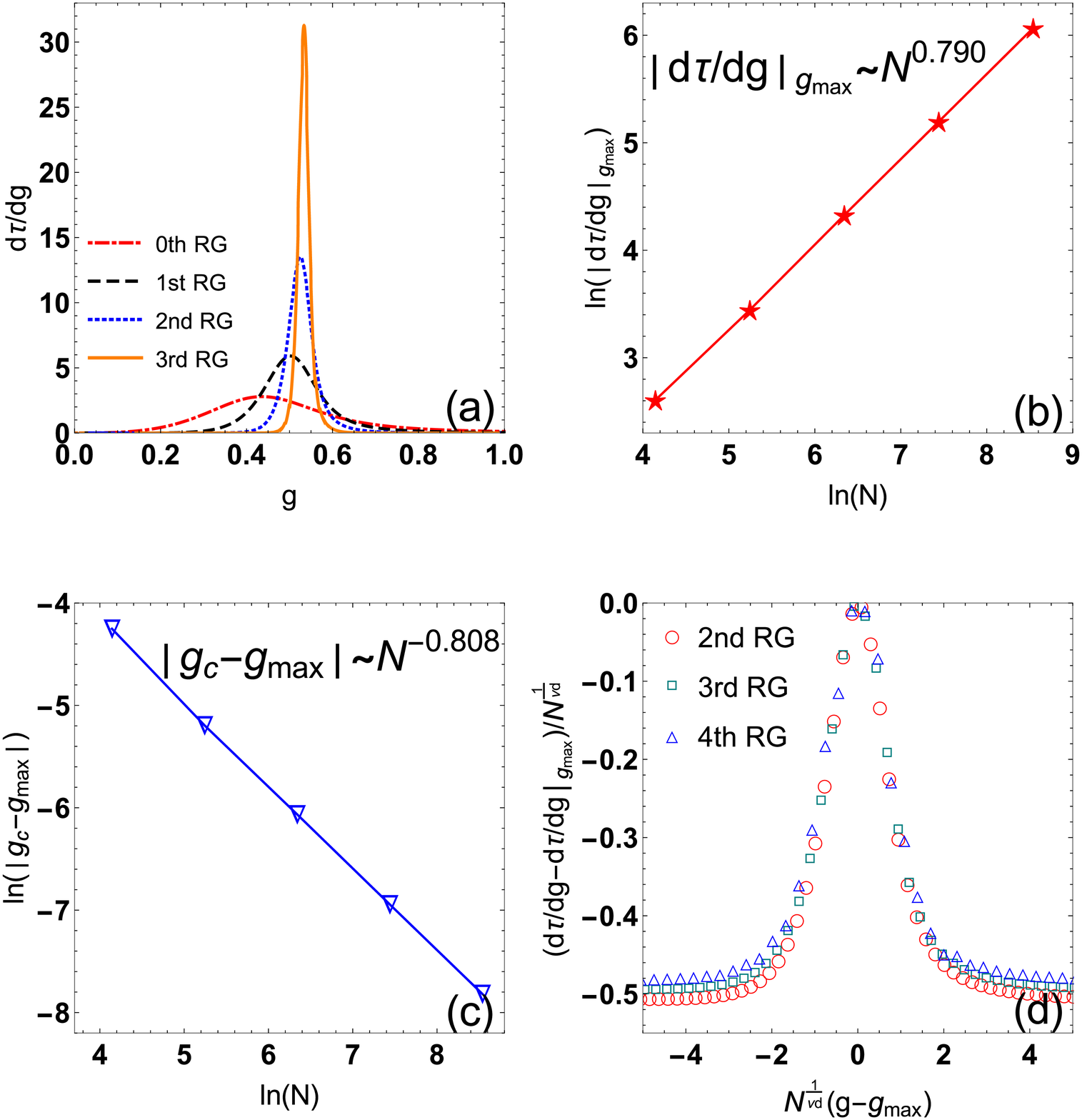}
 	\caption{(a) First derivative of multipartite entanglement $\mathrm{d}\tau/\mathrm{d}g$ versus $g$  for different RG iterations on the triangular lattice.
 		(b) The logarithm of the absolute value of maximum $\ln\left| \mathrm{d} \tau/\mathrm{d}g\right| $ versus the  logarithm of the triangular lattice size $\ln(N)$, which is linear and shows a scaling behavior. (c)The scaling behavior of $g_\mathrm{max}$ in terms of system size $N$ for the triangular lattice, where $g_\mathrm{max}$ is the position of the maximum derivative of multipartite entanglement. (d) The finite-size scaling through renormalization treatment with the  correlation length exponent $\nu=0.63$ for the multipartite entanglement. The curves  corresponding to different system sizes approximately collapse onto a single one  for this triangular lattice.}
 	\label{figure3}
 \end{figure}

Based on Eqs. \ref{RG equation1} and \ref{seven-partite}, we  compute the seven-partite entanglement between seven sites in the basic cluster   which can represent  different system sizes after completing the corresponding RG iterations.
The seven-partite entanglement  $\tau$ as a function of $g$ for different RG iterations on the triangular lattice is plotted in Fig. \ref{figure2}. It can be observed that  these curves  cross
each other at the critical point  and two different saturated values of multipartite entanglement associated with two  phases: the ferromagnetic  phase ($g>g_\mathrm{c}^\mathrm{t}$) and the paramagnetic phase ($g<g_\mathrm{c}^\mathrm{t}$) are developed. As the size of  system becomes large, the two phases are diverged more clearly. In particular, the saturated value of multipartite entanglement in the ferromagnetic  phase approaches the maximum. Here, in order to
 provide a possible physical explanation, we display   the  bipartite entanglements of formation $E_{\rm{f}}(\rho_{12})$  and $E_{\rm{f}}(\rho_{1|2\cdots7})$  as functions of  coupling strength $g$ for the zeroth RG iteration  in the inset of Fig. \ref{figure2}. The bipartite entanglement $E_{\rm{f}}(\rho_{12})$ first increases from zero to the maximum and then decreases to zero monotonically, while another one $E_{\rm{f}}(\rho_{1|2\cdots7})$ increases monotonically with $g$ until the saturated value is arrived.  As the coupling strength grows from zero,  the increased  probability for the spin pair staying at the entangled state
 leads to the generation of bipartite entanglement.  Only when the competition between the interaction and quantum fluctuation reaches a counterbalance at the critical point, the bipartite entanglement $E_{\rm{f}}(\rho_{12})$  reaches its maximum \cite{gu2003entanglement}, agreeing with the results of Ref. \cite{PhysRevA.95}. Then the  exchange couplings play a dominant role and
 keep the system staying at the ferromagnetic phase, which results in the decrease of bipartite entanglement  between two neighboring sites.  We can find  the reason why the multipartite entanglement approaches the maximum in the  ferromagnetic phase  according to Eq. \ref{multipartite entanglement}. On the one hand, the increase of coupling strength brings the value of  quantum  entanglement $E_{\rm{f}}(\rho_{1|2\cdots7})$ up to the maximum.  On the other hand, the bipartite entanglements  between two neighboring sites  are very little since the strong coupling strength. It can be concluded that the multipartite entanglement is not maximal at the critical point as the bipartite entanglement be, and the multipartite entanglement  is richer in the  ferromagnetic phase than the paramagnetic phase.

 Although above interpretation is superficial, it provides inspiration for further understanding. At zero field the model exhibits  ferromagnetic behavior with net magnetization in the $z$ direction. The ground state is twofold degenerate and a product state with spins pointing in the  $z$ direction, i.e. $|+\rangle = |\uparrow \uparrow\cdots\rangle$ or $|-\rangle = | \downarrow\downarrow \cdots \rangle$. In the large-field limit, ground state is also a product state with all  spins being polarized to the direction of field. Although  no bipartite entanglement exists  in both  cases,  there is another possible solution for the ground state at zero field, namely, the superposition of the
 degenerate states, which may be a  Greenberger-Horne-Zeilinger (GHZ)-like state $|\rm{G}\rangle = 1/\sqrt{2}(|+\rangle + |-\rangle)$ with genuine multipartite entanglement \cite{Montakhab2010}. Therefore, the multipartite entanglement approaches the maximum in the  ferromagnetic phase, and these results may provide  us  a further insight in the entanglement distribution and QPT for many-body systems.

More information on the location and the order of the QPT can be obtained by consideration of the derivatives of the multipartite entanglement with respect to the coupling strength. We plot the derivatives of multipartie entanglement $\mathrm{d}\tau/\mathrm{d}g$ as a function of $g$ for different RG iterations in Fig. \ref{figure3}(a). It can be seen from Fig. \ref{figure3}(a)  that the first derivative of the multipartie entanglement exhibits a nonanalytic behavior, which indicates that the QPT of this system is a second-order QPT. The scaling behavior of the maximum of $\mathrm{d}\tau/\mathrm{d}g$ versus $N$ is displayed in the Fig. \ref{figure3}(b), which is a linear behavior of $\ln(|\mathrm{d}\tau/\mathrm{d}g|_{g_\mathrm{max}})$ versus $\ln(N)$. Based on  numerical analysis, we can obtain  $|\mathrm{d}\tau/\mathrm{d}g|_{g_\mathrm{max}}\sim N^{\mu'_1}$ where the critical exponent $\mu'_1 \simeq 0.790$. It has been found that   the correlation length exponent is the inverse of critical exponent in the one-dimensional spin chain systems \cite{kargarian2008the,ma2011quantum}. For this triangular lattice, the relation between the correlation length exponent and critical exponent has a new form.   The correlation length exponent $\nu$ gives the divergent behavior of correlation length in the vicinity of $g_\mathrm{c}$, i.e., $\xi \sim |g-g_\mathrm{c}|^{-\nu}$.  Under the RG transformations, the correlation length scales as $\xi \rightarrow \xi_n=\xi/\lambda^n $, where  $\lambda= \sqrt{3}$ is the scale of the length of  the side for  each RG iteration and related to $N$, i.e. $7\times (\lambda^n)^d = N$.
For the $n$th RG iteration, the renormalized coupling strength $g_n$ is still a function of original one $g$. Since $ \xi_n \sim |g_n-g_\mathrm{c}|^{-\nu}$ and $\left|\mathrm{d}\tau/\mathrm{d}g|_{g_\mathrm{max}}\right| \sim \left| \mathrm{d}g_n/\mathrm{d}g|_{g_\mathrm{c}}\right|$, we can derive that
\begin{equation}
\left| \frac{\mathrm{d}\tau}{\mathrm{d}g} \right|_{g_\mathrm{max}} \sim N^{\frac{1}{\nu d}}.
\end{equation}
Comparing with $|\mathrm{d}\tau/\mathrm{d}g|_{g_\mathrm{max}}\sim N^{\mu_1}$,  the relation of  critical exponent and correlation length exponent is obtained, namely $\mu_1=1/(\nu d)$. Furthermore, the value of coupling strength $g_\mathrm{max}$ corresponding to the maximum of $\mathrm{d}\tau/\mathrm{d}g$ for each RG iteration gradually tends toward the critical point $g_\mathrm{c}$, which  indicates another scaling behavior displayed in Fig. \ref{figure3}(c), i.e., $|g_\mathrm{c} -g_\mathrm{max}| \sim N^{-\mu'_2}$ where the critical exponent $\mu'_2\simeq 0.808$.  This  critical exponent is also related to the correlation length exponent $\nu$ in the vicinity of the critical point.  The scaling of the position of maximum $g_\mathrm{max}$ comes from the behavior of the correlation length $\xi$ near the critical point.  As the thermodynamic limit is approached, the
correlation length $\xi \sim N^{1/d}$. Comparing with $\xi \sim |g - g_{c}|^{-\nu}$, the scaling form $|g_\mathrm{c} - g_\mathrm{max} | \sim N^{-1/(\nu d)}$ can be obtained, which implies that $\mu_2=1/(\nu d)$.  We can observe that the second critical exponent is  in good agreement with the first one $\mu_1=\mu_2=1/(\nu_\mathrm{t} d) \simeq 0.794$ where $\nu_\mathrm{t}$ is the  correlation length exponent for the triangular lattice as shown in Eq. \ref{correlation length exponent}.
It is noted that the above relation can  also be proved by the numerical results $\mu'_1 \simeq 0.790$ and $\mu'_2\simeq 0.808$.  Based on the  divergence of derivative of multipartite entanglement, we plot
$(\mathrm{d}\tau/\mathrm{d}g-\mathrm{d}\tau/\mathrm{d}g|_{g_\mathrm{max}})/N^{\frac{1}{\nu d}}$ versus $N^{\frac{1}{\nu d}}(g-g_\mathrm{max})$ for different RG iterations in Fig. \ref{figure3}(d). These curves for
different  system sizes  approximately collapse onto a single one, which is a manifestation of the existence of finite-size scaling for the multipartite entanglement \cite{Nature2002Scaling,Kargarian2007}. We can conclude that the multipartite entanglement is a good indicator to signify the criticality of the  transverse-field quantum Ising model on the triangular  lattice.

\subsection{Sierpi\'nski fractal lattice}

\begin{figure}[ht]
	\centering
	\subfigure{ \label{figure4a} \includegraphics[width=7cm]{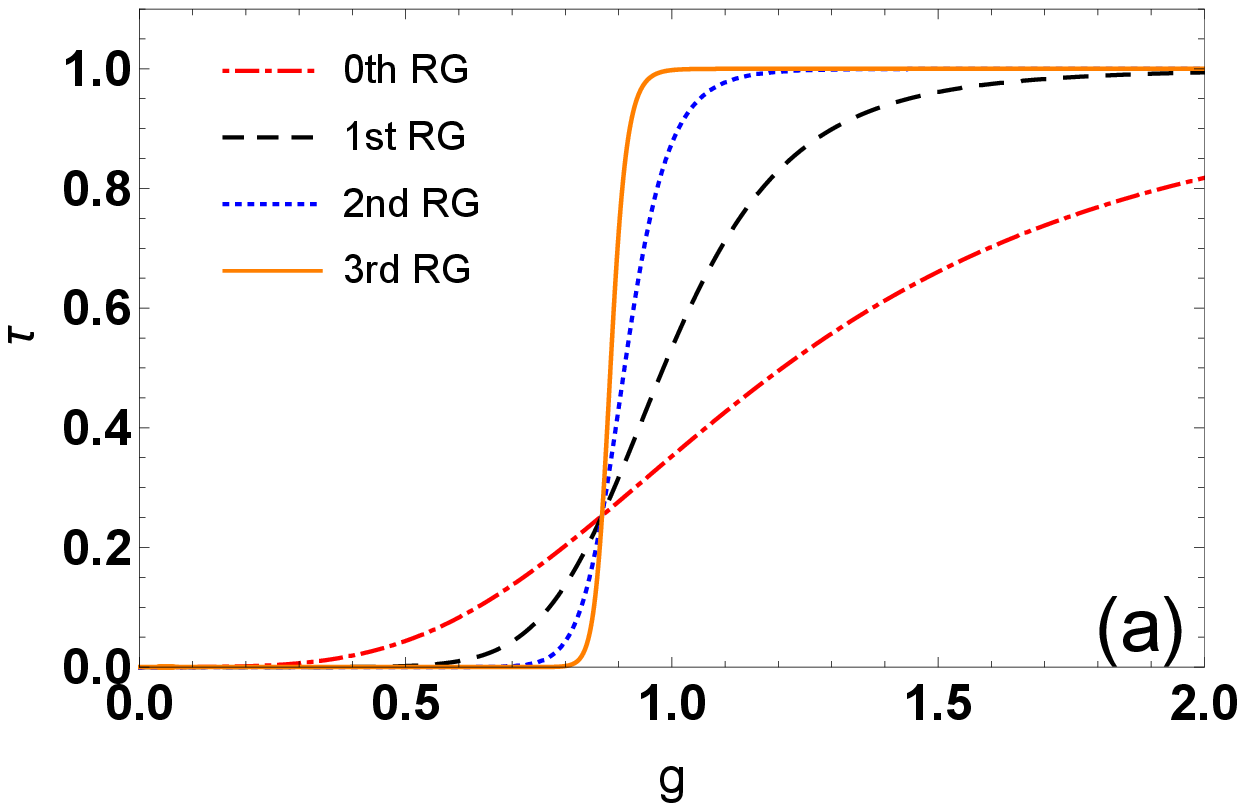}}
	\subfigure{ \label{figure4b} \includegraphics[width=7cm]{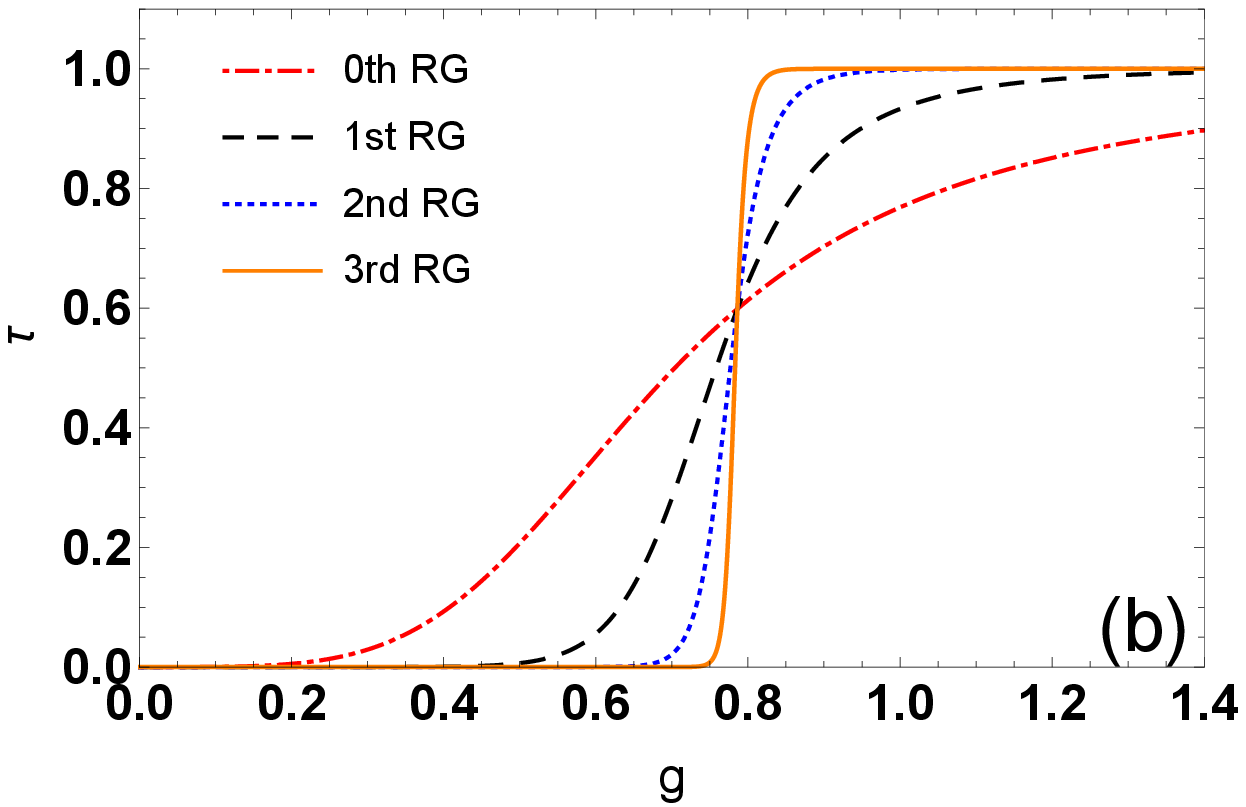}}
	\caption{The evolutions of multipartite entanglement $\tau$ versus $g$ for different RG iterations on the (a) Sierpi\'nski triangular lattice and (b) Sierpi\'nski pyramid lattice.}
	\label{figure4}
\end{figure}

 \begin{figure}[htpb]
	\centering \includegraphics[width=8cm]{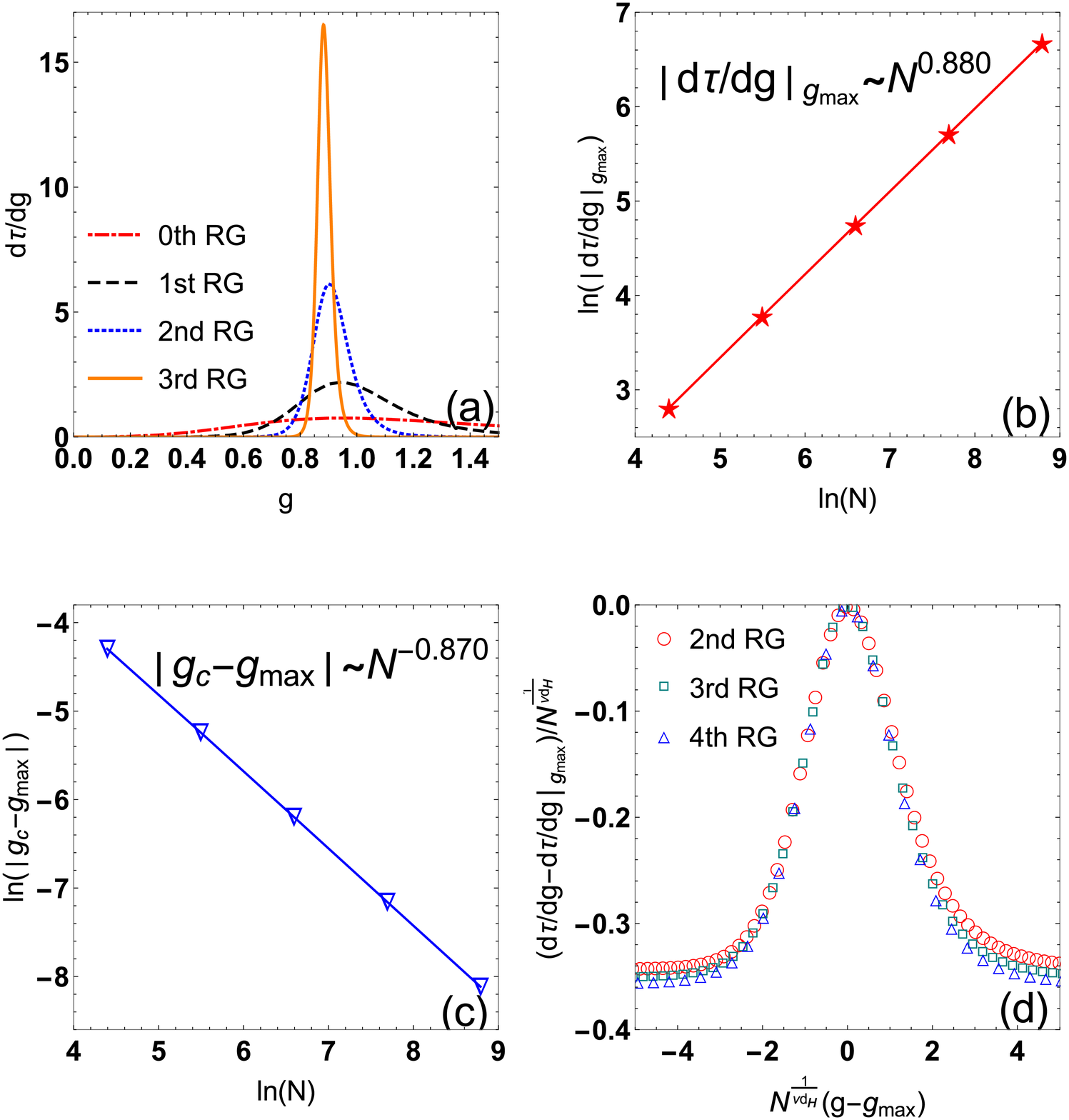}
	\caption{(a) The first derivative of multipartite entanglement $\mathrm{d}\tau/\mathrm{d}g$ versus $g$  for different RG iterations on the Sierpi\'nski triangular lattice.
		(b) The logarithm of the absolute value of maximum $\ln\left| \mathrm{d}\tau/\mathrm{d}g\right| $versus the  logarithm of the Sierpi\'nski triangular lattice size  $\ln(N)$, which is linear and shows a scaling behavior. (c) The scaling behavior of $g_\mathrm{max}$ in terms of system size $N$ for the Sierpi\'nski triangular lattice, where $g_\mathrm{max}$ is the position of the maximum derivative of multipartite entanglement.  (d) The finite-size scaling through renormalization treatment with the  correlation length critical exponent $\nu=0.720$ for the multipartite entanglement. The curves corresponding to different system sizes approximately collapse onto a single one  for this Sierpi\'nski triangular lattice.}
	\label{figure5}
\end{figure}
 Next, we consider the transverse-field Ising model on the  fractal lattices which are the generalizations of the Sierpi\'nski pyramid in $\kappa$ spatial dimensions. For $\kappa=2$ and $\kappa=3$, the  fractal lattices are Sierpi\'nski triangle and
pyramid lattices, respectively, as depicted in Fig. \ref{figure1b} and  Fig. \ref{figure1c}, whose  Hausdorff  dimensions can be calculated by $d_\mathrm{H} =\log(\kappa+1)/\log2$. Here, we choose the  site  labeled by $1$ as the \textquotedblleft node\textquotedblright,
and according to the numbers of sites in basic clusters of fractal lattices, we investigate the tripartite entanglement for Sierpi\'nski triangle lattice and four-partite entanglement for Sierpi\'nski pyramid lattice, respectively.  The renormalized tripartite and four-partite entanglements can be obtained from  Eqs. \ref{RG equation2} and \ref{multipartite entanglement}.
The results about multipartite entanglement versus the reduced coupling strength $g$ for different RG iterations on the Sierpi\'nski triangular and pyramid lattices are displayed in Fig. \ref{figure4}.  It is clearly observed from Fig. \ref{figure4}  that the evolutions of multipartite entanglement on the fractal lattices are similar to that on the triangular lattice. As the size of the Sierpi\'nski triangular  lattice increases, the tripartite entanglement produces two different  saturated values that corresponding to  two different phases. The paramagnetic order at $g<g_\mathrm{c}^\mathrm{St}$ induces the quantum fluctuation and leads to the destruction of tripartite entanglement. In contrast, as  $g$ becomes large, the  ferromagnetic order gradually  builds the tripartite entanglement. The performance of four-partite entanglement in the Sierpi\'nski pyramid lattice is analogous to the tripartite entanglement in the Sierpi\'nski triangle lattice, one obvious difference is the positions of  intersection  points since the critical points of these two fractal lattices are diverse. These two figures reveal that as the thermodynamic limit is touched  by increasing the RG iterations, the multipartite entanglement can be used to detect the critical points of the  fractal lattices.
\begin{figure}[t]
	\centering
 \includegraphics[width=8cm]{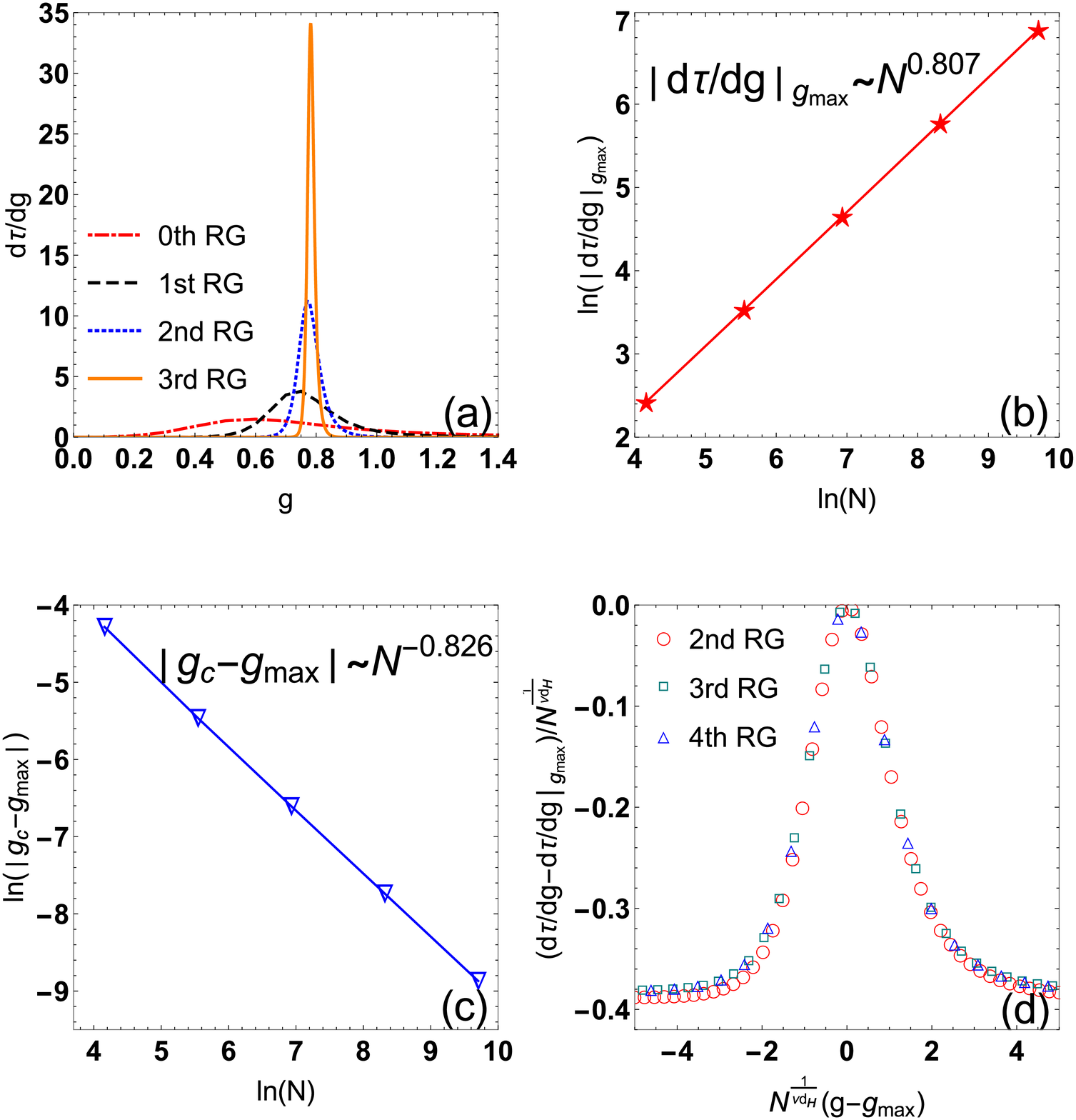}
	\caption{(a) The first derivative of multipartite entanglement $\mathrm{d}\tau/\mathrm{d}g$ versus $g$  for different RG iteration on the Sierpi\'nski  pyramid lattice.
		(b) The logarithm of the absolute value of maximum $\ln\left| \mathrm{d}\tau/\mathrm{d}g\right| $versus the  logarithm of the Sierpi\'nski pyramid lattice size  $\ln(N)$, which is linear and shows a scaling behavior. (c) The scaling behavior of $g_\mathrm{max}$ in terms of system size $N$ for the Sierpi\'nski pyramid lattice, where $g_\mathrm{max}$ is the position of the maximum derivative of multipartite entanglement. (d) The finite-size scaling through renormalization treatment with the  correlation length critical exponent $\nu=0.617$ for the multipartite entanglement. The curves  corresponding to different system sizes approximately collapse onto a single one  for this Sierpi\'nski pyramid lattice.}
	\label{figure6}
\end{figure}

The appearance of nonanalytic behavior in some quantity, often accompanied by a scaling behavior, is a feature of second-order QPT. The  nonanalytic phenomenons of the first derivative of multipartite entanglement  near the critical point and the scaling behaviors  for the two-dimensional triangular lattice have been shown in Fig. \ref{figure3}. In the following, we pay our attention to the cases of fractal lattices. The first derivatives of multipartite entanglement
 $\mathrm{d}\tau/\mathrm{d}g$ versus $g$  for different RG iterations on the Sierpi\'nski triangular and pyramid lattices have been displayed in Fig. \ref{figure5}(a) and Fig. \ref{figure6}(a), respectively. The  nonanalytic behaviors of multipartite entanglement near the critical points become more prominent when the sizes of systems increase, which means that the first derivatives of multipartite entanglement are singular  near the critical points, and the systems both undergo the second-order QPTs. To further understand the relation between the renormlized multipartite entanglement and QPTs, we  explore the finite-size scaling behaviors of multipartite entanglement close to the critical points. The linear behaviors of $\ln(\left| \mathrm{d}\tau/\mathrm{d}g\right|_{g_\mathrm{max}})$ versus $\ln(N)$ are revealed in Figs. \ref{figure5}(b) and \ref{figure6}(b).  Numerical analysis confirmed that the maximum of $\mathrm{d}\tau/\mathrm{d}g$ obeys the following
finite-size scaling behavior: $\left| \mathrm{d}\tau/\mathrm{d}g|_{g_\mathrm{max}}\right| \sim N^{\mu'}$, where the critical exponent for Sierpi\'nski triangular lattice is $\mu'_3 \simeq 0.880$ as shown in Fig. \ref{figure5}(b)
and the one for Sierpi\'nski pyramid lattice is $\mu'_5 \simeq 0.807$ as shown in Fig. \ref{figure6}(b).   The correlation length exhibits exponential behavior near the critical point $g_\mathrm{c}$, i.e., $\xi \sim |g-g_\mathrm{c}|^{-\nu}$.  After the $n$th iteration, the correlation length scales as $\xi_n=\xi/\lambda^n \sim |g_n-g_\mathrm{c}|^{-\nu} $ with  $\lambda_{\mathrm{f}}= 2$.   $N$ and $\lambda_{\mathrm{f}}$ in the fractal lattice have the relation $N=N_0 \lambda_{\mathrm{f}}^{nd_\mathrm{H}}$, where $N_0 =3$ for $\kappa=2$ and $N_0 =4$ for $\kappa=3$. Since $\left|\mathrm{d}\tau/\mathrm{d}g|_{g_\mathrm{max}}\right| \sim \left| \mathrm{d}g_n/\mathrm{d}g|_{g_\mathrm{c}}\right|$, we can derive that
\begin{equation}
\left| \frac{\mathrm{d}\tau}{\mathrm{d}g} \right|_{g_\mathrm{max}} \sim N^{\frac{1}{\nu d_\mathrm{H}}}.
\end{equation}
Comparing with $|\mathrm{d}\tau/\mathrm{d}g|_{g_\mathrm{max}}\sim N^{\mu}$,  the relation of  critical exponent and correlation length exponent can be obtained $\mu=1/(\nu d_\mathrm{H})$ . It means that $\mu_3=1/(\nu_\mathrm{St} d_\mathrm{H}^\mathrm{St}) \simeq 0.876$  for Sierpi\'nski triangular lattice and $\mu_5=1/(\nu_\mathrm{Sp} d_\mathrm{H}^\mathrm{Sp}) \simeq 0.810$  for Sierpi\'nski pyramid lattice. The numerical results  listed above $\mu'_3 \simeq 0.880$ and  $\mu'_5 \simeq 0.807$  are in good agreement with the analytical results, i.e.  $\mu'_3 \simeq \mu_3$ and $\mu'_5 \simeq \mu_5$.

Furthermore,  the value of $g_\mathrm{max}$ corresponding to
the maximum of $\mathrm{d}\tau/\mathrm{d}g$ for each RG iteration gradually tends toward the critical
point $g_\mathrm{c}$. It indicates $|g_\mathrm{c} -g_\mathrm{max}| \sim N^{-\mu}$,  where the critical exponent for Sierpi\'nski triangular lattice is $\mu'_4 \simeq 0.870$ as shown in Fig. \ref{figure5}(c)
and the one for Sierpi\'nski pyramid lattice is $\mu'_6 \simeq 0.826$ as shown in Fig. \ref{figure6}(c).  These critical exponents $\mu_4$ and $\mu_6$ are directly related the correlation length exponents in the vicinities of the critical points. The correlation length is related to the  size of the system  in the thermodynamic limit, i.e., $\xi \sim N^{1/d_\mathrm{H}}$. Since $\xi \sim |g - g_{c}|^{-\nu}$, then the scaling form $|g_\mathrm{c} - g_\mathrm{max} | \sim N^{-1/(\nu d_\mathrm{H})}$ can be obtained, which implies that $\mu=1/(\nu d_\mathrm{H})$. That is to say, for the Sierpi\'nski triangular lattice, $\mu_4=1/(\nu_\mathrm{St} d_\mathrm{H}^\mathrm{St}) \simeq 0.876$  and for the Sierpi\'nski pyramid lattice, $\mu_6=1/(\nu_\mathrm{Sp} d_\mathrm{H}^\mathrm{Sp}) \simeq 0.810$. The numerical results are also consistent with the analytical results, i.e.,  $\mu'_4 \simeq \mu_4$ and $\mu'_6 \simeq \mu_6$.

Finally,   it is possible to make all the data from different RG iterations collapse onto a single curve by choosing a suitable scaling function and taking into account the distance of the
maximum of the derivatives of multipartite entanglement from the critical point \cite{Nature2002Scaling,Kargarian2007}. We display
$(\mathrm{d}\tau/\mathrm{d}g-\mathrm{d}\tau/\mathrm{d}g|_{g_\mathrm{max}})/N^{\frac{1}{\nu d_\mathrm{H}}}$ versus $N^{\frac{1}{\nu d_\mathrm{H}}}(g-g_\mathrm{max})$ for different RG iterations on the Sierpi\'nski triangular lattice in Fig. \ref{figure5}(d) and on the Sierpi\'nski pyramid lattice in Fig. \ref{figure6}(d).  These curves  approximately collapse onto a single  universal one, which  is a manifestation of the existence of finite-size scaling for the multipartite entanglement. These results justify that the RG implementation of multipartite entanglement truly capture the critical behaviors of the transverse-field quantum Ising model on the fractal lattices.

It is well known that the cornerstone
of the theory of critical phenomena is the  universality, which indicates that the critical behavior is depend on the dimension of system and the symmetry of  chosen order parameter \cite{Sachdev1999,Nature2002Scaling}.  From above discussions, it can be confirmed that the critical behaviors of these lattices  depend on their dimensions. Furthermore, we want to point out that the multipartite entanglement may be a better choice than bipartite entanglement or quantum correlations for studying the many-body systems.  On the one hand, the bipartite entanglement has limited ability to capture the  characters of the many-body systems.  A typical example is the $N$-qubit  GHZ state $\left| G \right\rangle = 1/\sqrt{2}( \left| \uparrow\uparrow \cdots\uparrow\right\rangle +\left| \downarrow\downarrow \cdots\downarrow \right\rangle)$, which has been proved to be an $N$-partite entangled state.  However, its reduced density matrix of two spins ($i$ and $j$) $ \rho_{ij} = 1/2 ( \left| \uparrow\uparrow \right\rangle \left\langle \uparrow\uparrow\right| +\left| \downarrow\downarrow \right\rangle \left\langle \downarrow \downarrow \right|)$  is a separable mixed state and has no entanglement. In Fig. \ref{figure2},  the bipartite entanglement only reaches the maximum near the critical point, while the multipartite entanglement reaches the maximum in a more extensive region.  The multipartite entanglement in ferromagnetic phase may be a valuable resource for the quantum information processing tasks.  We may lose this important information  and  have less chance to know the entanglement distribution of the many-body system  if we only consider the bipartite entanglement.  On the other hand, although the bipartite entanglement has been
successfully proved to capture the quantum critical points  of some many-body systems, it has
been indicated that the bipartite entanglement  may fail to characterize the real quantum critical points \cite{Osborne2002,Qian2005,Yang2005}. For
example, the concurrence may show no special behavior at the real critical point of
the one-dimensional frustrated spin-1/2 Heisenberg model \cite{Qian2005}. In this sense, the multipartite entanglement
provides a global view and more physical insights into the characters of many-body systems
and may have some advantages over bipartite entanglement or quantum correlation for studying the  many-body systems \cite{Montakhab2010,Hofmann2013}.

\section{\label{section3}Quantum coherence and Quantum phase transitions in lattices}

In this section, we  choose the quantum coherence as an indicator to  study the  QPTs in  the transverse-field quantum Ising model on the
triangular lattice and fractal lattices by using  the QRG method. It is noted that  the existence of QPT is independent of the chosen physical quantity.
In order to quantify the amount of quantum coherence, the $l_1$-norm and quantum relative entropy coherence have been proposed in Ref. \cite{PRLcoherence}.
 Besides, some other effective quantifiers of quantum
coherence, such as the quantum coherence based on the trace distance and quantum Jensen-Shannon divergence (QJSD) have been put forward in the later works \cite{Shao2015The,Rana2016Trace,Radhakrishnan2016}. Here, we choose the quantum coherence based on the QJSD \cite{Radhakrishnan2016} to study the QPTs of lattices.  The QJSD is a measure of distinguishability between two quantum states \cite{PhysRevA.72.052310}
\begin{equation}
J(\rho,\sigma)=S\left(\frac{\rho+\sigma}{2}\right)-\frac{S(\rho)+S(\sigma)}{2},
\end{equation}
where $S(\rho)=-\rm{Tr}\rho \log_2 \rho $ is the von Neumann entropy. In Ref. \cite{PhysRevA.77.052311}, the   metric character of QJSD has been discussed   and  a true metric based on the square root of QJSD has been proposed as follows
 \begin{equation}
  D(\rho,\sigma)=\sqrt{J(\rho,\sigma)}.
  \end{equation}
 It is noted  that this metric  verifies the triangle inequality in addition to satisfying the distance axioms, and it is  a valuable tool since its metric properties. Moreover, it has been proven to be true for qubit  and  qudit systems \cite{Bri2008Properties,Lin2017Spectral,Radhakrishnan2016}, and the measure of quantum coherence based on the square root of the QJSD is given by
  \begin{equation}
  C(\rho)=\sqrt{S\left(\frac{\rho+\rho_{\rm{dia}}}{2}\right)-\frac{S(\rho)+S(\rho_{\rm{dia}})}{2}}
  \end{equation}
  where $\rho_{\rm{dia}}$ is the incoherent state obtained from $\rho$ by deleting all off-diagonal elements \cite{PRLcoherence}. Quantum coherence  are usually ascribed to the off-diagonal elements of a density matrix with respect to a reference basis.   We fix the computational basis $\left\lbrace |0\rangle, |1\rangle \right\rbrace $ as the reference basis, where $ |0\rangle$ and $ |1\rangle$ are the eigenvectors of spin operator $\sigma^z$.

\subsection{ triangular lattice}
\begin{figure}[ht]
	\centering
	\includegraphics[width=7cm]{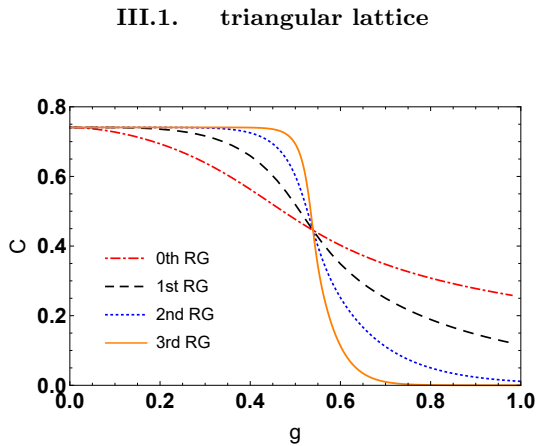}
	\caption{The evolution of quantum coherence $C$ versus $g$ for different RG iterations on the triangular lattice.}
	\label{figure7}
\end{figure}
First, we investigate the quantum coherence between two nearest-neighbor sites on  the  triangular lattice. As shown in Fig. \ref{figure1a}, the basic cluster contains seven sites, for simplicity, we apply the quantum coherence $C(\rho_{1 2})$ between sites $1$ and $2$ to study the  QPT,  where the reduced density  matrix $\rho_{12}$ can be obtained by  tracing over the sites $3,\cdots,7$.
The   quantum coherence  $C$ as a function of $g$ for different RG iterations on the triangular lattice is plotted in Fig. \ref{figure7}. By comparing Fig. \ref{figure2}, it is clear that  two different saturated values of quantum coherence are developed, however, the behaviors of quantum coherence for two different phases are completely opposite to the ones of multipartite entanglement.
This is due to the fact that the quantum coherence is basis-dependent. We choose the eigenvectors of $\sigma^z$ as the reference basis and call it $\sigma^z$-basis. When the  coupling strength $g$ is small enough,  the  external field  induces the  quantum fluctuation and lead to all the spins being polarized along the direction of the field, i.e. the $x$ axis. It means that the $\sigma^x$ terms contribute to the off-diagonal elements of density matrix, which leads the generation of quantum coherence in the $\sigma^z$-basis \cite{PhysRevA.96.012341}. Then a saturated value of quantum coherence is reached in the thermodynamic limit. As $g$ increases, the  exchange coupling gradually plays a dominant  role and keeps the system  staying at the ferromagnetic phase,
the contribution from the $\sigma^x$ terms for quantum coherence almost disappears. Therefore, the quantum coherence tends to zero in ferromagnetic phase after enough RG iterations.

\begin{figure}[ht]
 	\centering
 		 \includegraphics[width=8cm]{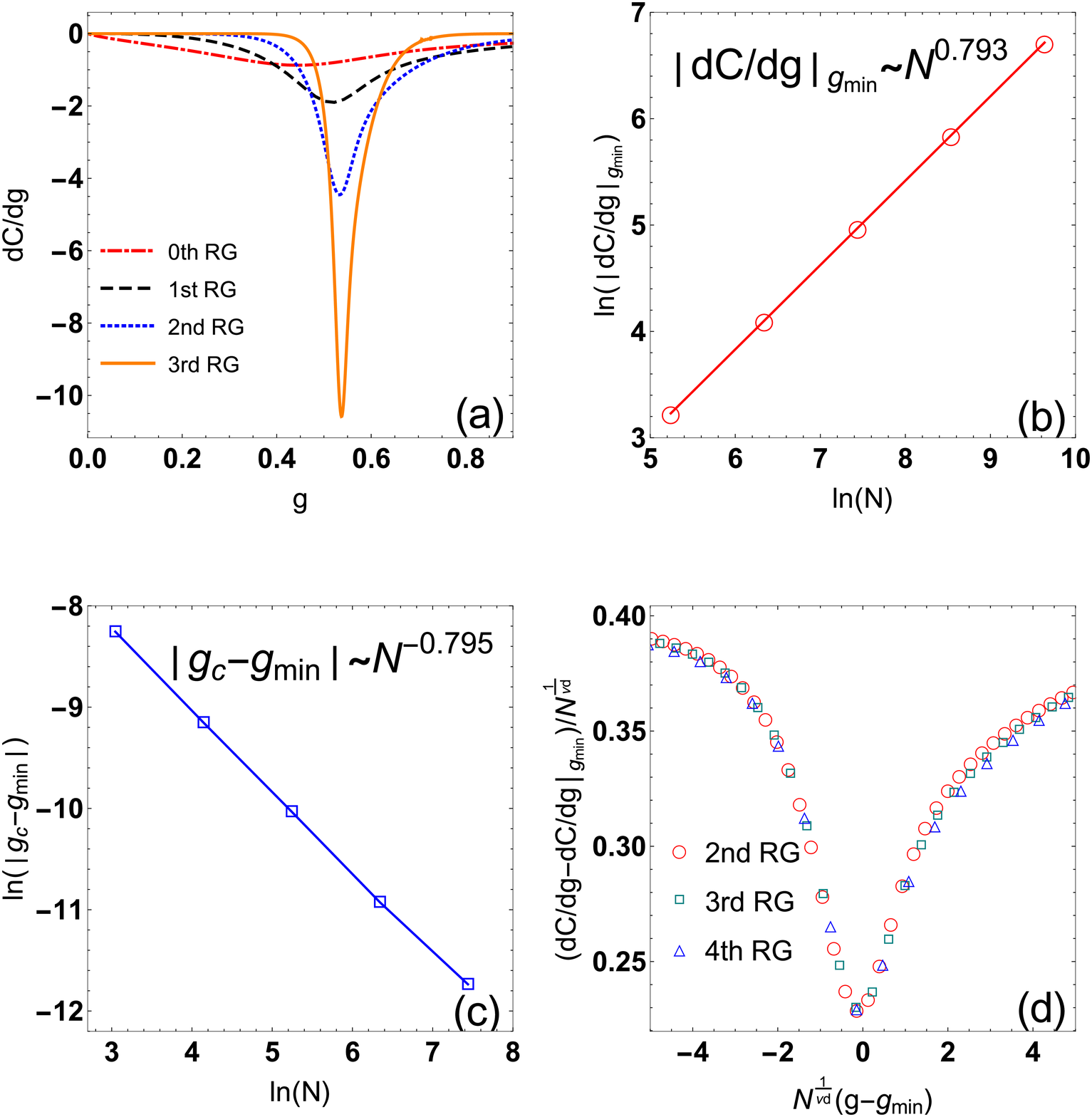}
 	\caption{(a) First derivative of quantum coherence $\mathrm{d}C/\mathrm{d}g$ versus $g$  for different RG iteration on the triangular lattice.
 		(b)The logarithm of the absolute value of minimum $\ln\left| \mathrm{d}C/\mathrm{d}g\right| $ versus the  logarithm of the triangular lattice size $\ln(N)$, which is linear and shows a scaling behavior. (c) The scaling behavior of $g_\mathrm{min}$ in terms of system size $N$ for the triangular lattice, where $g_\mathrm{min}$ is the position of the minimum derivative of quantum coherence. (d)
 		 The $(\mathrm{d}C/\mathrm{d}g-\mathrm{d}C/\mathrm{d}g|_{g_\mathrm{min}})/N^{1/(\nu d)}$ versus $N^{1/(\nu d)}(g-g_\mathrm{min})$ for different RG iterations on  triangular lattice where the  correlation length critical exponent $\nu=0.630$. The curves corresponding to different system sizes approximately collapse onto a single one.}
 \label{figure8}
 \end{figure}
In order to obtain the precise location of critical point and  the order of QPT, we look at the derivatives of the quantum coherence $\mathrm{d}C/\mathrm{d}g$ as a function of $g$ for different RG iterations in Fig. \ref{figure8}(a). It is quite clear from Fig. \ref{figure8}(a) that the first derivative of the quantum coherence exhibits a nonanalytic behavior in the vicinity of the critical point,  which is a feature of the second-order QPT.
The position of the minimum of $\mathrm{d}C/\mathrm{d}g$ is gradually close to the critical point as the size of system increases.  A  linear behavior of $\ln(|\mathrm{d}C/\mathrm{d}g|_{g_\mathrm{min}})$ versus $\ln(N)$ is displayed in Fig. \ref{figure8}(b). The critical exponent $\mu''_1$ for this scaling behavior is $|\mathrm{d}C/\mathrm{d}g|_{g_\mathrm{min}}\sim N^{\mu''_1}$ where $\mu''_1 \simeq 0.793$.   Another scaling behavior is shown in Fig. \ref{figure8}(c), i.e.,  $|g_\mathrm{c} -g_\mathrm{min}| \sim N^{-\mu''_2}$ where the critical exponent $\mu''_2\simeq 0.795$.  Using the similar analysis in the previous section, we can obtain
the relation between the correlation length exponent and critical exponent for the  triangular lattice, namely, $\mu_1=\mu_2=1/(d\nu_\mathrm{t}) \simeq 0.794$, which indicates the numerical result is  consistent with the analytical one, i.e.  $\mu''_1 \simeq \mu''_2 \simeq \mu_1$.  In Fig. \ref{figure8}(d), we plot
$(\mathrm{d}C/\mathrm{d}g-\mathrm{d}C/\mathrm{d}g|_{g_\mathrm{min}})/N^{\frac{1}{\nu d}}$ versus $N^{\frac{1}{\nu d}}(g-g_\mathrm{min})$ for different RG iterations.  All the data from different $N$ collapse onto a single curve, which  provides a manifestation of the existence of finite-size scaling for the quantum coherence. It can be concluded that the quantum coherence is a good indicator to signify the criticality of the  transverse-field quantum Ising model on the triangular lattice.

The multipartite entanglement and quantum coherence
are both able to capture the characteristics of ground state and show special
behaviors at the real critical point of the triangular lattice, which lead to the appearance
of similar properties. At the same time, these similar properties enable us to obtain
consistent critical exponents of multipartite entanglement and quantum coherence, i.e. $\mu'_1 \simeq \mu'_2 \simeq \mu''_1 \simeq \mu''_2$, which is the presentation of the universality of QPT and also demonstrates that the existence
of QPT is independent of the chosen physical quantity.

\begin{figure}[ht]
	\centering
	\subfigure{ \label{figure9a} \includegraphics[width=7cm]{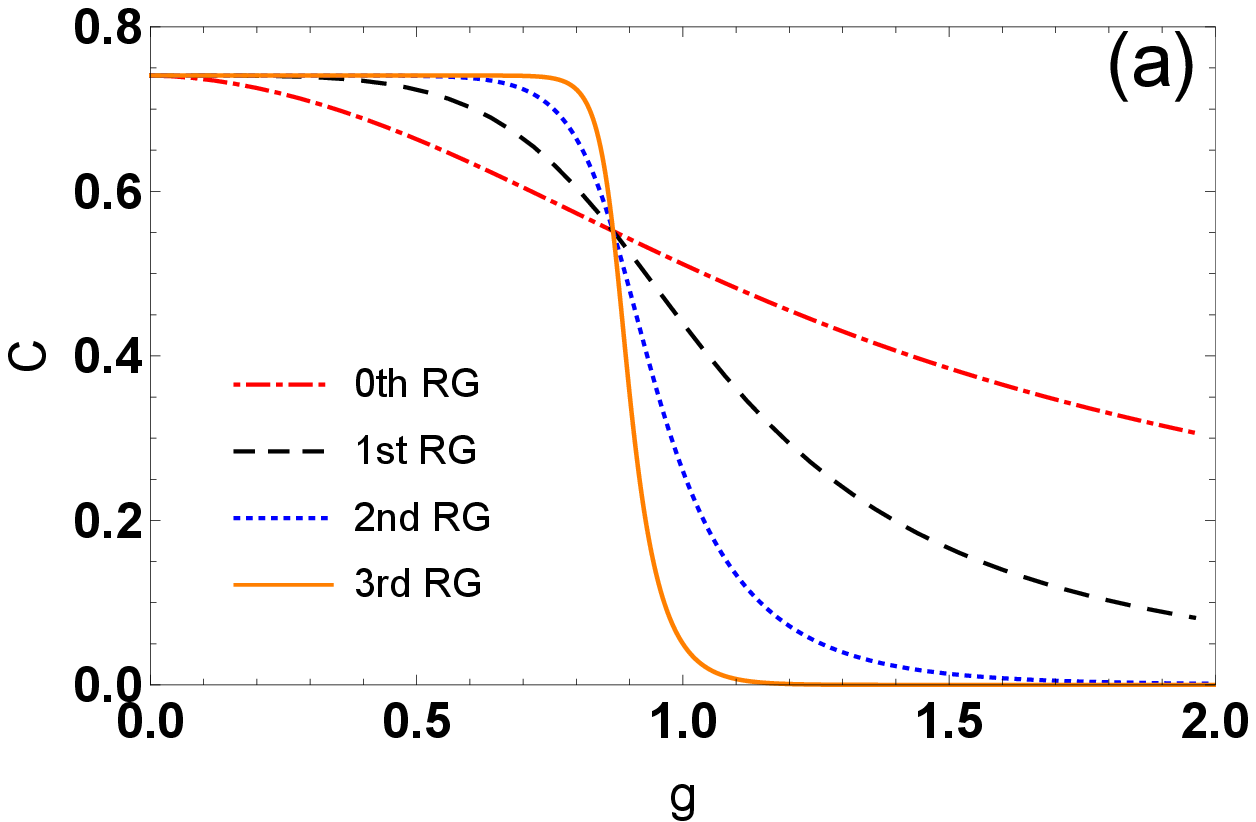}}
	\subfigure{ \label{figure9b} \includegraphics[width=7cm]{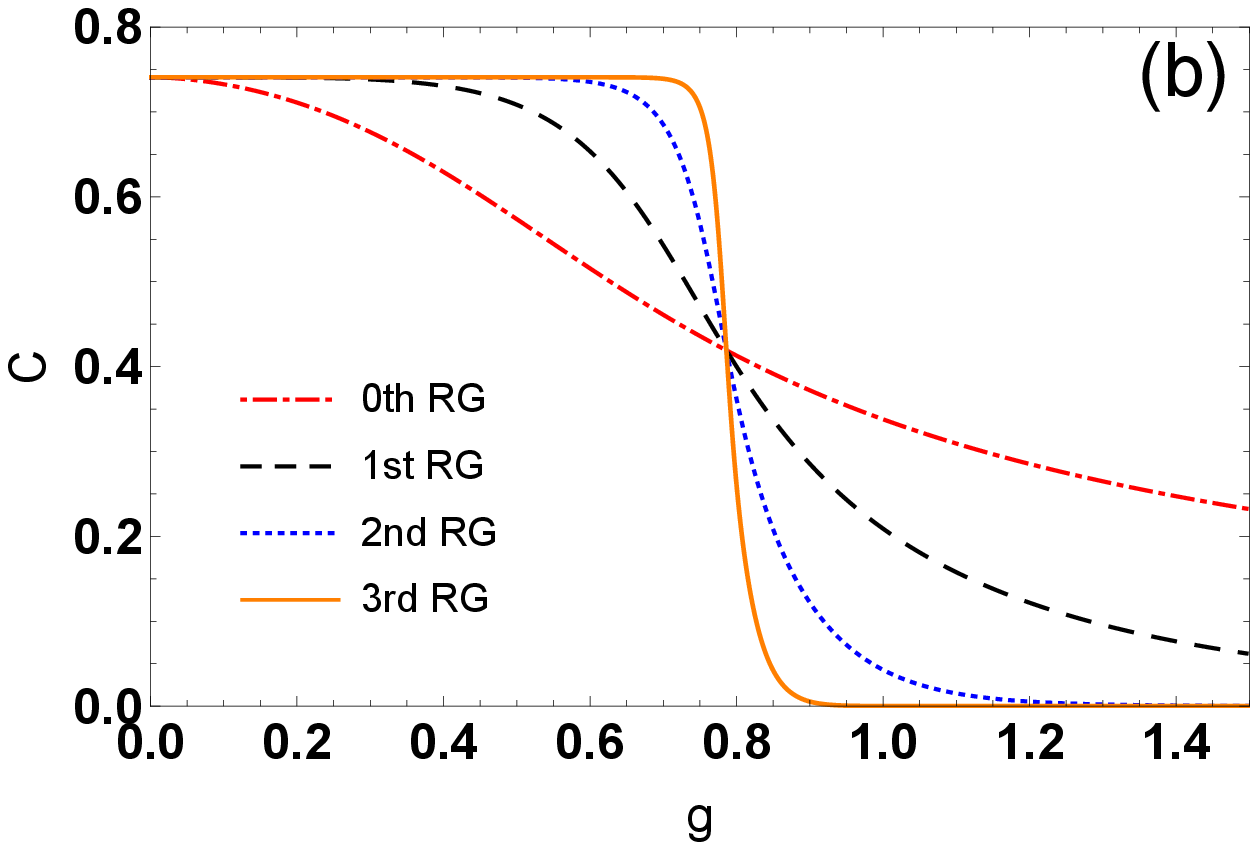}}
	\caption{The evolution of quantum coherence $C$ versus $g$ for different RG iteration on the (a) Sierpi\'nski triangular lattice and (b) Sierpi\'nski pyramid lattice.}
	\label{figure9}
\end{figure}
\begin{figure}[htpb]
	\centering \includegraphics[width=8cm]{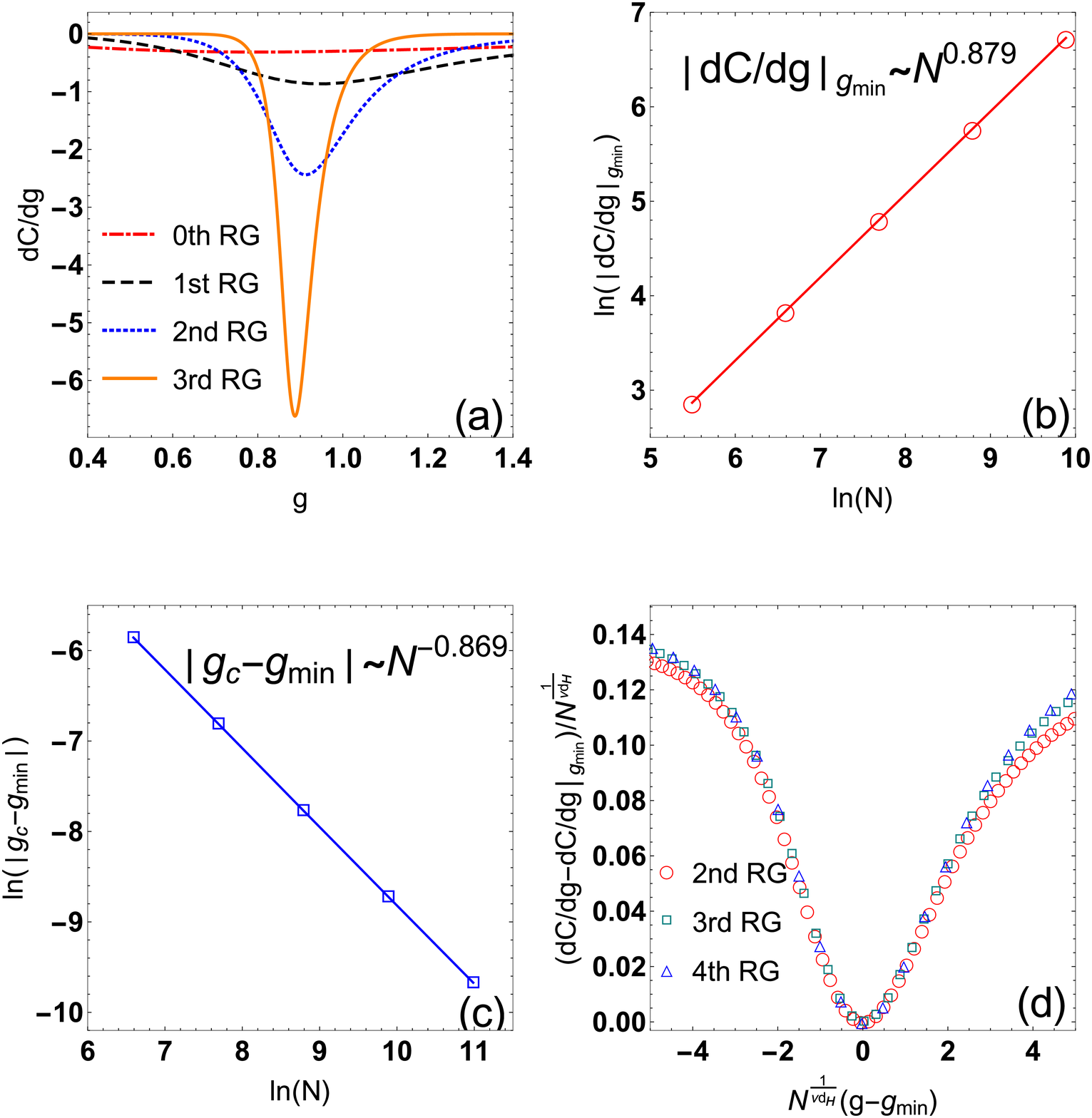}
	\caption{(a) The first derivative of quantum coherence $\mathrm{d}C/\mathrm{d}g$ versus $g$  for different RG iterations on the Sierpi\'nski triangular lattice.
	(b) The logarithm of the absolute value of minimum $\ln\left| \mathrm{d}C/\mathrm{d}g\right| $ versus the  logarithm of the Sierpi\'nski triangular lattice size $\ln(N)$, which is linear and shows a scaling behavior.
	(c) The scaling behavior of $g_\mathrm{min}$ in terms of system size $N$ for the Sierpi\'nski triangular lattice, where $g_\mathrm{min}$ is the position of the minimum derivative of quantum coherence.
	(d) The $(\mathrm{d}C/\mathrm{d}g-\mathrm{d}C/\mathrm{d}g|_{g_\mathrm{min}})/N^{1/(\nu d_\mathrm{H})}$ versus $N^{1/(\nu d_\mathrm{H})}(g-g_\mathrm{min})$ for different RG iterations on Sierpi\'nski triangular lattice where the  correlation length critical exponent $\nu=0.720$. The curves  corresponding to different system sizes approximately collapse onto a single one.}
	\label{figure10}
\end{figure}

\subsection{Sierpi\'nski fractal lattice}

Next, we turn to the transverse-field quantum Ising model on the  fractal lattices and focus on the quantum coherence  between two nearest-neighbor sites which are labeled by $1$ and $2$ in Figs. \ref{figure1b} and \ref{figure1c}. The results about quantum coherence $C$ versus the reduced coupling strength $g$ for different RG iterations on the Sierpi\'nski triangular and pyramid lattices are shown in Fig. \ref{figure9}.
It is found that the evolutions of quantum coherence on the fractal lattices are similar to that on the triangular lattice.
 As the size of the  lattice increases, the quantum coherence produces two different  saturated values that corresponding to  two different phases.  Quantum coherence   is stronger in the paramagnetic phase than the  ferromagnetic phase for both of the fractal lattices.
 The positions of  intersection  points are different since the critical points of these two fractal lattices are not same. As the thermodynamic limit is touched by increasing the RG iterations, the quantum coherence can be used to detect the critical points of the transverse-field quantum Ising model on fractal lattices.
\begin{figure}[htpb]
	\centering \includegraphics[width=8cm]{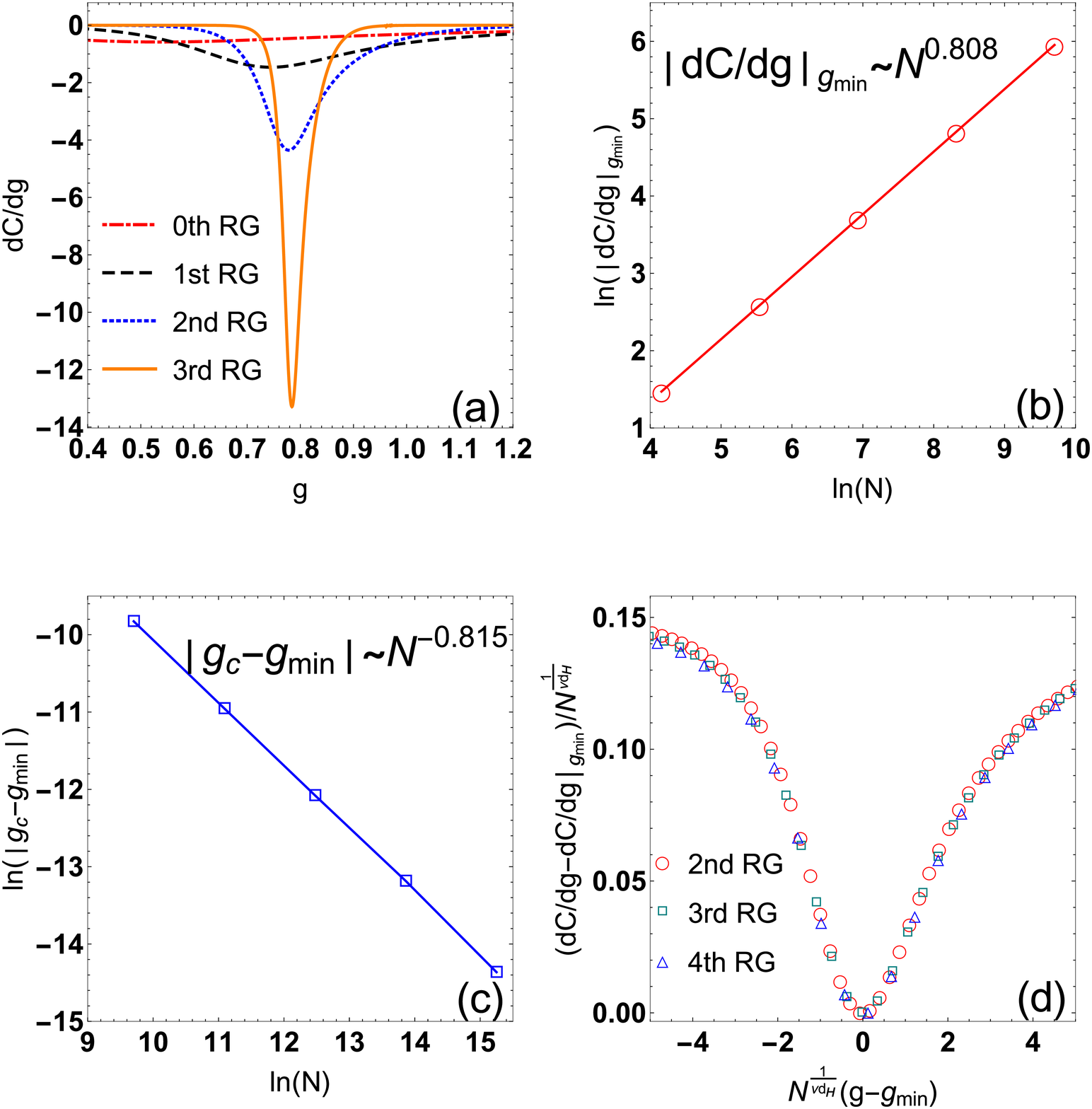}
	\caption{(a) The first derivative of quantum coherence $\mathrm{d}C/\mathrm{d}g$ versus $g$  for different RG iteration on the Sierpi\'nski  pyramid lattice.
		(b) The logarithm of the absolute value of minimum $\ln\left| \mathrm{d}C/\mathrm{d}g\right| $ versus the  logarithm of the Sierpi\'nski pyramid lattice size $\ln(N)$, which is linear and shows a scaling behavior. (c) The scaling behavior of $g_\mathrm{min}$ in terms of system size $N$ for the Sierpi\'nski pyramid lattice, where $g_\mathrm{min}$ is the position of the minimum derivative of quantum coherence. (d) The $(\mathrm{d}C/\mathrm{d}g-\mathrm{d}C/\mathrm{d}g|_{g_\mathrm{min}})/N^{1/(\nu d_\mathrm{H})}$ versus $N^{1/(\nu d_\mathrm{H})}(g-g_\mathrm{min})$ for different RG iterations on Sierpi\'nski pyramid lattice where the  correlation length critical exponent $\nu=0.617$. The curves corresponding to different system sizes approximately collapse onto a single one.}
	 \label{figure11}
\end{figure}

The nonanalytic features of the first derivatives of quantum coherence at the critical points of the Sierpi\'nski triangular and pyramid lattices  are given in Figs. \ref{figure10}(a) and  \ref{figure11}(a), respectively.  Two systems both  exhibit singular properties  as the increase of RG iterations. We also explore the finite-size scaling behaviors of renormalized quantum coherence close to the critical points of fractal lattices. The linear behaviors of $\ln(\left| \mathrm{d}C/\mathrm{d}g\right|_{g_\mathrm{min}})$ versus $\ln(N)$ are revealed in Figs. \ref{figure10}(b) and \ref{figure11}(b). The result of numerical analysis
confirms that the  minimum of $\mathrm{d}C/\mathrm{d}g$ obeys the following
finite-size scaling behavior: $\left| \mathrm{d}C/\mathrm{d}g|_{g_\mathrm{min}}\right| \sim N^{\mu''}$, where the critical exponent for Sierpi\'nski triangular lattice is $\mu''_3 \simeq 0.879$ as shown in Fig. \ref{figure10}(b)
and the one for Sierpi\'nski pyramid lattice is $\mu''_5 \simeq 0.808$ as shown in Fig. \ref{figure11}(b).  Figs \ref{figure10}(c) and \ref{figure11}(c) present the results of our analysis for another kind of scaling behavior of $g_{\rm{min}}$ in terms of system
size $N$,  $|g_\mathrm{c} -g_\mathrm{min}| \sim N^{-\mu''}$,  where  $\mu''_4 \simeq 0.869$ for Sierpi\'nski triangular lattice and $\mu''_6 \simeq 0.815$  for Sierpi\'nski pyramid lattice.  The relations between  correlation length exponents and critical exponents of quantum coherence can be analytically  obtained as well.  In the case of Sierpi\'nski triangular lattice, $\mu_3=\mu_4=1/(\nu_\mathrm{St} d_\mathrm{H}^\mathrm{St}) \simeq 0.876$, and in the case of  for Sierpi\'nski pyramid lattice,  $\mu_5=\mu_6=1/(\nu_\mathrm{Sp} d_\mathrm{H}^\mathrm{Sp}) \simeq 0.810$.  The numerical results
are in agreement with the analytical ones.
It can be seen from Figs. \ref{figure10}(d) and \ref{figure11}(d) that the curves  corresponding to different sizes of system  clearly collapse on a single universal curve. These results justify that the RG implementation of quantum coherence truly capture the critical behaviors of the transverse-field quantum Ising model on the fractal lattices.

We want to emphasize that the study of  quantum coherence on fractal lattices provides us more insights into the characteristics of the  fractal lattices. First, the Hausdorff  dimensions  determining the  relations between  correlation length exponents and critical exponents of quantum coherence are confirmed again. Second, the critical exponents of  quantum coherence are consistent with the ones of multipartite entanglement, namely $\mu''_3 \simeq \mu''_4 \simeq \mu'_3 \simeq \mu'_4$ for the Sierpi\'nski triangular lattice and $\mu''_5 \simeq \mu''_6 \simeq \mu'_5 \simeq \mu'_6$ for the Sierpi\'nski pyramid lattice. These results not only demonstrate that the existence
of QPT is independent of the chosen physical quantity, but also are the indication of the universality of QPT.

\begin{figure}[ht]
	\centering
	\includegraphics[width=4cm]{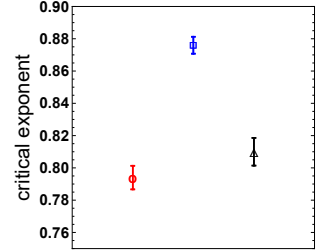}
	\caption{The critical exponents of the triangular lattice (red circle), Sierpi\'nski  triangular lattice (blue square) and Sierpi\'nski pyramid lattice (black triangle) with error bars  denoting the simulation errors.}
	\label{figure12}
\end{figure}

At last, based on the obtained analytical and numerical results of the critical exponents, we display  the critical exponents of the triangular lattice and Sierpi\'nski fractral lattices  with error bars   denoting the simulation errors in Fig. \ref{figure12}.  The
error bars originate from the standard deviation and indicate the errors between the numerical and  analytical results. It can be confirmed again that  the numerical  values of critical exponents obtained from multipartite entanglement and quantum coherence are consistent with the analytical ones.
\section{\label{section4}Conclusions}

We investigate the performances of multipartite entanglement and quantum coherence in the quantum phase transitions for transverse-field quantum Ising model on the triangular lattice and Sierpi\'nski  fractal  lattices by employing the QRG method.  It is shown that the quantum criticalities of these high-dimensional models closely relate to the behaviors of the multipartite entanglement and quantum coherence.
As the thermodynamic limit is approached, the multipartite entanglement and quantum coherence for these models both develop two different values corresponding to two phases, i.e., the ferromagnetic phase and paramagnetic phase. However, the performances of the multipartite entanglement and quantum coherence in two phases are completely different. The multipartite entanglement in ferromagnetic phase  is richer than the one in paramagnetic phase since  the GHZ-like state is more  likely to exist in the ferromagnetic phase.
Nevertheless, the quantum coherence in  paramagnetic phase  is stronger than the one in  ferromagnetic phase, for the reason that  the external field may induce the  quantum fluctuation and lead to some  spin being polarized along the direction of the field, and  the $\sigma^x$ terms contribute to the off-diagonal elements of density matrix, which leads the generation of quantum coherence.

 Moreover, the singularities and  finite-size scaling behaviors for each lattice  can be obtained by calculating the first derivatives of multipartite entanglement and quantum coherence.
Although a similar analysis has been performed using the bipartite entanglement in Ref. \cite{PhysRevA.95}, the authors have only  given one kind of scaling behavior for each lattice.  In contrast,  we have obtained all the scaling behaviors as  mentioned in  Ref. \cite{Nature2002Scaling},  especially the ones which characterize how the critical points $g_c$ of these models are touched as the increase of system sizes.
The critical exponents are  related to the correlation length exponents and dimensions of lattices, which is due to the fact that the universality
of quantum phase transition is dependent on the effective dimension.
The multipartite entanglement and  quantum coherence are both able to capture the characteristics of ground states and show special
behaviors at the critical points, which lead to the appearance
of similar properties and  consistent critical exponents. It is the presentation of the universality of QPT and also demonstrates that the existence of QPT is independent of the chosen physical quantity.

 In general, the  multipartite entanglement and quantum coherence are  both good indicators
to detect the quantum phase transitions in the triangular lattice  and Sierpi\'nski fractal lattices.  We expect our results to be
of interest for a wide range of applications in  other high-dimensional lattices with help of the QRG method.

\section{Acknowledgments}

This project was supported by the National Natural Science Foundation of China (Grant No.11274274) and the Fundamental Research Funds for the Central Universities (Grant No.2017FZA3005 and 2016XZZX002-01).
%

\end{document}